\documentclass{aa}
\usepackage{graphicx}
\def\ssim{\setbox0=\hbox{$\propto$}%
\setbox1=\hbox{$<$}\dimen0=\ht1%
\advance\dimen0by-1.2pt\,\lower.6\dimen0%
\copy0\kern-\wd0\raise.4\dimen0\copy1 \,}

\def\gsim{\setbox0=\hbox{$\propto$}%
\setbox1=\hbox{$>$}\dimen0=\ht1%
\advance\dimen0by-1.2pt\,\lower.6\dimen0%
\copy0\kern-\wd0\raise.4\dimen0\copy1\,}

\def\lambdab{\lambda\mkern-9mu\lower1.2pt\hbox{$\mathchar'26$}}%

\begin{document}
   \title{Stellar evolution with rotation VII: }

\subtitle{ Low metallicity models 
and the blue to red supergiant ratio in the SMC}

 \author{A. Maeder, G. Meynet}

     \institute{Geneva Observatory CH--1290 Sauverny, Switzerland\\
              email: Andre.Maeder@obs.unige.ch\\
              email:  Georges.Meynet@obs.unige.ch      }

   \date{Received 15 March 2001 / Accepted 12 April 2001}

\abstract{ We calculate a grid of models with and without
the effects of axial rotation for massive stars in the range of 9 
to  60 M$_{\odot}$ and metallicity $Z$ = 0.004 appropriate for  the 
SMC. Remarkably, the ratios $\Omega/\Omega_{\mathrm{crit}}$ 
of the angular velocity to the break--up angular velocity
grow strongly during the evolution of high mass stars, 
contrary to the situation at $Z$ = 0.020.
The reason is that at low $Z$, mass loss is smaller and the removal
of angular momentum during evolution  much weaker, also there is an
 efficient outward transport of angular momentum 
by meridional circulation. Thus, a much larger 
fraction of the stars at lower $Z$ reach
break--up velocities and  rotation may thus be
a dominant effect at low
$Z$.  The models with rotation well
account for the long standing problem of the large numbers
of red supergiants observed in low $Z$ galaxies, while current
 models with mass loss were predicting no red supergiants.
We discuss in detail the physical effects of rotation which 
favour a redwards evolution in the HR diagram. The models also predict 
large N enrichments during the evolution of 
  high mass stars. The predicted relative N--enrichments
are larger at  $Z$ lower than  solar and this is
in very good agreement  with the observations for A--type supergiants
in the SMC.
               }

   \maketitle
%

\section{Introduction}

The aim of this work is to study the effects of rotation in
massive stars at low metallicity  $Z$ as in the SMC. There are 3 main
problems emerging in this context. 1.-- Firstly, we want
to examine whether there  are differences between the effects
of rotation at low $Z$ and at solar metallicity. Surprisingly,
 we find some striking differences, which suggest that rotation may 
be a much more important effect in the evolution of stars at $Z$ 
lower than solar. 2.-- Then,
we want to address the problem of the number ratio $B/R$
of blue to red supergiants, which
is one of the most severe problems in
 stellar evolution. As stated
by Kippenhahn and Weigert (\cite{KW90}) about stars on the blue
loops in the He--burning phase, the present phase
`` ... is a sort of magnifying glass, revealing relentlessly the 
faults of calculations of earlier phases.'' As long as 
the problem  is not
solved, we cannot  correctly  predict the populations of massive stars
in galaxies at various metallicities $Z$, neither the spectral evolution of 
galaxies, nor the chemical yields. 3.-- A third important question
concerns the chemical abundances of massive stars at low metallicities.
In particular, we recall that very  large nitrogen enrichments
have been found for the A--type supergiants in the SMC (Venn
\cite{Venn98, Venn99}).  There, the relative $N/H$ excesses 
can reach a factor of 10, much larger than in the Galaxy. This
is a strong constraint, which needs to be examined carefully.  

Let us comment a bit more on the  embarassing
 problem of the number ratio 
 $B/R$, which  has been reviewed by Langer \& Maeder
(\cite{LM95}) and by Maeder \& Meynet (\cite{MMARAA}). To make a long 
story short, the observations of nearby galaxies show
that there are much more red supergiants at metallicities lower 
than solar, so that 
$B/R$ decreases steeply with decreasing
 metallicity (Humphreys \& McElroy \cite{Hum84}). 
Large differences  in the $B/R$ ratio also exist between clusters 
in the Galaxy and in the SMC, where the red supergiants 
 were found more numerous by an order of a magnitude
 (Meylan \& Maeder \cite{Meylan82}).

There are no
sets of models which correctly predict the observed trend of decreasing 
$B/R$ with decreasing $Z$, as emphasized by Langer \& Maeder
(\cite{LM95}). All kind of models were examined: models
with Schwarzschild's criterion (cf. Stothers \& Chin  \cite{Stoth92}),
models with Schwarzschild's criterion and overshooting
(cf. Schaller et al. \cite{Schall92}), models with Ledoux criterion
(cf. Stothers \& Chin \cite{Stoth92}; Brocato \& Castellani 
\cite{Broc93}), models with semiconvection (cf. Arnett 
\cite{Arn91}), models with semiconvective diffusion (cf. 
Langer \& Maeder \cite{LM95}). The comparisons generally show 
that the models with Schwarzschild's criterion (with or without
overshooting) may reproduce the observed $B/R$ at solar metallicity, 
while they fail at lower $Z$.  At the other extreme, the models with
the Ledoux criterion and those with semiconvection  reproduce  well
$B/R$ at the metallicities of the SMC, but they fail at higher $Z$.
This shows that, at least in part, the problem is related to
the size of the core and to the mixing  efficiency outside the core.

Of course, other effects such as mass loss, convection,
opacities and  metallicities
 play a role. In particular, as evidenced by the Geneva
grids of models (Meynet et al. \cite{Mey94}; cf. also Maeder 
\cite{M81}) 
a growth of mass loss favours the  formation 
of more red supergiants in the He-burning phase.
However, the $B/R$ problem in the SMC cannot be solved by  mass 
loss, because the mass loss rates at lower $Z$, as in the SMC, are
smaller than in the Galaxy and this produces fewer
red supergiants.

In some recent works, it has been shown that the account of the
 axial rotation of stars
changes all  the model outputs; tracks in the HR diagram, lifetimes,
surface abundances, etc... (Meynet \& Maeder \cite{MMV}; Heger \& 
Langer \cite{HL00}). Noticeably rotation, both by its
effects on  internal mixing and on mass loss 
was found to favour the
redward motions.
 This is a positive  indication and we now further
explore it. A grid of stellar models with rotation
in the range of 9   to 60 M$_{\odot}$ and 
metallicity $Z$ = 0.004 was  constructed.
It will provide a basis for comparison with the SMC observations.

In Sect. 2, we  briefly discuss the
improvements brought in the model physics. In Sect. 3, we 
discuss the evolution of the internal rotation law $\Omega(r)$
and in Sect. 4 of the surface rotation velocities $v$.
The models with zero rotation are  briefly discussed in Sect. 5.
In Sect. 6, the results for the HR diagram and  the lifetimes 
are shown. Sect. 7 is devoted to the study of 
the $B/R$ ratio from the numerical models; we also account for
the results in terms of the physical properties of stellar models.
In Sect. 8, we analyze the chemical abundances at the stellar
surface and compare the results to observations of
A--type supergiants in the SMC.
Sect. 9 gives the conclusions.

\section{Physics of the models}

We apply the treatment of the  hydrostatic effects 
appropriate to differential
rotation, as described by 
Meynet \& Maeder (\cite{MMI}). We recall that the classical
 treatment frequently applied is as a matter of fact not correct
in the presence of shellular
differential rotation. We also account  for the  
rotational distortion and for
the von Zeipel theorem. The  $T_{\mathrm{eff}}$
given here corresponds to the value  as observed
 for an average orientation
angle between the axis of rotation and the direction of the observer.
We  include the effects of  shear diffusion 
(Zahn \cite{Za92}; Maeder \cite{MII}) and  
the effects of the  meridional
circulation, as studied  by Maeder \& Zahn (\cite{MZIII}).
The same physics of the models  as 
described in  
 Meynet \& Maeder (\cite{MMV}, paper V) is used here, 
with a few  improvements, which are mentioned below.

\subsection{Meridional circulation, shears  and horizontal
turbulence}

Meridional circulation plays a major role in the redistribution
of the angular momentum in stars. Let us write
 the equations for the radial term of the vertical component
$U(r)$ of the meridional circulation 
(cf.  Maeder \& Zahn \cite{MZIII}), which will be essential for 
the following discussions,

\begin{eqnarray}
U(r) &=&  \frac{P}{\overline{\rho} \overline{g} C_{\!P} \overline{T}
\, [\nabla_{\rm ad} - \nabla +  (\varphi/\delta) \nabla_{\mu}] } \nonumber \\
& & \qquad\qquad \times   \left\{  \frac{L}{M_\star} (E_\Omega + E_\mu) + \!
\frac{C_P}{\delta} \frac{\partial \Theta }{\partial t} \! \right\} ,
\label{Ufinal}
\end{eqnarray}

\noindent
with the current notations as given in the quoted paper.
The main term in the braces in the second member
 is $E_{\Omega}$. If we ignore secondary terms,
 it behaves
essentially like,

 \begin{equation}
E_{\Omega}  \simeq  \frac{8}{3} \left[ 1 - \frac{{\Omega^2}}
{2\pi G\overline{\rho}}\right] \left( \frac{\Omega^2r^3}{GM}
\right) .
 \end{equation}

\noindent
The overlined expressions like $\overline{\rho}$ mean the average
over the considered equipotential.
The term with the minus sign in the square bracket is the 
Gratton--\"{O}pik term, which becomes important in the outer layers
due to the decrease of the local density; it
can produce negative values of $U(r)$. A  negative $U(r)$
means a circulation going down along the polar axis and up 
in the equatorial plane, thus making an outwards transport 
of angular momentum. The positive or negative values 
of $U(r)$ play a major role
in massive star evolution.

In paper V we did not account for the effects of
horizontal turbulence on the shears. This was done by
Talon \& Zahn (\cite{TaZa97}). They found
that the diffusion coefficient for the shears
is modified by the horizontal turbulence. The change can be 
an increase or a decrease of the diffusion coefficient depending 
on the various parameters, as discussed below. Thus,
 we  also  have to include the developments by 
Talon \& Zahn (\cite{TaZa97}), we have

\begin{eqnarray}
D =  \frac{ (K + D_{\mathrm{h}})}
{\left[\frac{\varphi}{\delta} 
\nabla_{\mu}(1+\frac{K}{D_{\mathrm{h}}})+ (\nabla_{\mathrm{ad}}
-\nabla_{\mathrm{rad}}) \right] }\; \times \\[2mm] \nonumber
 \frac{H_{\mathrm{p}}}{g \delta} \; 
\left [ \alpha\left( 0.8836\Omega{d\ln \Omega \over d\ln r} \right)^2
-4 (\nabla^{\prime}  -\nabla) \right]\\[4mm]
\mathrm{with} \; \;
 D_{\mathrm{h}} \simeq |r U (r)|
\end{eqnarray}

\noindent
where $D_{\mathrm{h}}$ is the coefficient of  horizontal diffusion 
(cf. Zahn \cite{Za92}).
We ignore here the thermal coupling effects
discussed by Maeder (\cite{MII})
because they were found to be relatively small
 and  they increase the
numerical complexity. Interestingly, we see that
in regions where $\nabla_{\mu} \simeq 0$, Eq. (3)  leads us to replace 
$K$ by  $(K+D_{\mathrm{h}})$ in the usual expression 
(cf. Talon \& Zahn \cite{TaZa97}), i.e. it reinforces slightly
the diffusion in regions which are close to chemical homogeneity.
On the contrary, in regions where $\nabla_{\mu}$ dominates
with respect to $(\nabla_{\mathrm{ad}} -\nabla_{\mathrm{rad}})$,
 the transport
is proportional to $D_{\mathrm{h}}$ rather than to $K$,
 which is quite logical since the diffusion is then
determined by  $D_{\mathrm{h}}$ rather than by thermal effects.
The above result shows the importance of the
treatment for the meridional circulation, since in turn it
determines the size of  $D_{\mathrm{h}}$ and to some extent
the diffusion by shears. The equation of transport of angular momentum
and of chemical elements are followed explicitly.

\subsection{Mass loss rates and their dependence on rotation}

For the mass loss rates, the recent data by
Kudritzki and Puls (\cite{KudrPuls00}) for OB stars and 
B--type supergiants are applied, they are
completed  for A--F supergiants
by the expressions by de Jager et al. (\cite{deJag88}). 
For the red supergiants, there are a number of recent
parametrizations proposed (cf. Willson \cite{Wills00}),
from low to high rates. We choose finally  the medium--high
rates as they were proposed by Vanbeveren et al. 
(\cite{Vanbe98}). For WR stars, the rates by Nugis \& Lamers 
(\cite{NuLa00}) were applied. These rates account for the clumping of
matter in the winds from WR stars and they are lower by about
a factor of 2 than those  used in the Geneva grids (cf. Schaller et al.
\cite{Schall92}). 

We have to account for rotation effects on the mass loss rates.
In our previous work (Meynet \& Maeder \cite{MMV}), we used
the expression derived by Langer (\cite{La98}),
based on the numerical models by Friend \&
Abbott (\cite{Fr86}). However, these models did not include the gravity
darkening predicted by the von Zeipel formula. A new expression has
been derived by the application of the wind theory over the surface
of a rotating star, taking also into account various improvements
in the model of a rotating star (Maeder \& Meynet \cite{MMVI}).
The ratio of the mass loss rate of a star rotating with an angular velocity
$\Omega$ to that of a non rotating star 
at the same location in the HR diagram behaves as

\begin{equation}
\frac{\dot{M} (\Omega)} {\dot{M} (0)} =
\frac{\left( 1  -\Gamma\right)
^{\frac{1}{\alpha} - 1}}
{\left[ 1 - \frac{\Omega^2}
{2 \pi G \rho_{\rm{m}}}-\Gamma \right]
^{\frac{1}{\alpha} - 1}}
\; .
\end{equation}

\noindent
The term $\Gamma$ is the usual Eddington factor for electron scattering
opacities and $\rho_{\rm{m}}$ is the average density internal
to the surface equipotential.
 The term $\frac{\Omega^2}{2 \pi G \rho_{\mathrm{m}}}$
is a function of the ratio of the rotational velocity to the 
break--up velocity

\begin{equation}
\frac{\Omega^2}{2 \pi G \rho_{\mathrm{m}}} \simeq
\frac{4}{9} \frac{v^2}{v_\mathrm{crit}^2} \; ,
\end{equation}

\noindent
with deviations of less than  $\sim 3 \%$ up to
$ \frac{v}{v_{\mathrm{crit}}}$ = 0.8. Here,
$v_\mathrm{crit} = \left(\frac{2}{3} 
\frac{GM}{R_{\mathrm{pb}}}\right)^\frac{1}{2}$
 and $R_{\mathrm{pb}}$ is the polar
radius at break-up ($R_{\mathrm{pb}}$ can be taken equal to
the radius of the non--rotating star). The coefficient $\alpha$
in Eq. (5)
is a force multiplier, which depends on $T_{\mathrm{eff}}$.
We use the empirical force multipliers derived by 
Lamers et al. (\cite{Lam95}): $\alpha =$ 0.52, 0.24, 0.17 and 
0.15, for $\log T_{\mathrm{eff}}  \geq$ 4.35, 
4.30, 4.00 and 3.90 respectively.

Of course, we have to account for the fact that the empirical values
for the mass loss rates 
used for non--rotating stars are based on 
stars covering the whole range of rotational velocities. Thus, we
must apply a reduction factor to the empirical rates to make
them correspond to the non rotating case. After convolution of
the effects described by Eq. (5) over the observed distribution
of rotational velocities, taking  also into account that the axes
of orientation are randomly distributed, we estimate that the
average correcting factor is equal to about 0.8, which has
to be applied to the mass loss rates for the Main--Sequence (MS)
OB stars. For post--MS stars, the mass loss rates are too
uncertain anyway, and the multiplying correction factor is almost unity,
so that we do not apply any correction.

\subsection{Initial composition, opacities, nuclear reactions}

For a given metallicity $Z$ (in mass fraction), we estimate the initial helium mass fraction
$Y$ from the relation $Y= Y_p \cdot \Delta Y/\Delta Z$  $Z$, 
where $Y_p$ is the primordial
helium abundance and $\Delta Y/\Delta Z$ the slope of 
the helium--to--metal enrichment law. Values of $Y_p$ between 0.23 and 0.24 are
given in the literature (see e.g. Pagel et al. \cite{Pagel};
 Izotov et al. \cite{Izotov};
 Peimbert et al. \cite{Peimbert}).
Observations of HII regions in blue compact dwarf
galaxies indicate values for $\Delta Y/\Delta Z$ equal
 to 1.7$\pm$0.9 (Izotov et al. \cite{Izotov}),
2.3$\pm$1.0 (Thuan \& Izotov \cite{Thuan}).
 Such values are in agreement with the $\Delta Y/\Delta Z$
value of 2.3$\pm$1.5 derived by Fernandes et al. 
(\cite{Fernandes}) from the study of nearby visual binary stars.
Here we set $Y_p$ = 0.23 and $\Delta Y/\Delta Z$ = 2.25
as in the recent grids of stellar models
by Girardi et al. (\cite{Girardi}).  
For the metallicity $Z$ = 0.004 considered in this work, we thus obtain
$X$ = 0.757 and $Y$ = 0.239.
For the mixture of the heavy elements, 
we adopted the same mixture as the one
used to compute the opacity tables. Opacities are from
Iglesias \& Rogers (\cite{Iglesias})
 complemented at low temperatures with the molecular opacities 
of Alexander (http://web.physics.twsu.edu/alex/wwwdra.htm).

The rates for the charged particle reactions are taken from  
the new NACRE compilation (Angulo et al., \cite{Angulo}). 
For the reaction $^{12}$C$(\alpha, \gamma)^{16}$O and for the 
range of temperatures corresponding to the He--burning phase,
the NACRE rate is a factor two higher
than the rate of Caughlan \& Fowler (\cite{Caughlan88})
 and amounts to 
about 80\% of the rate adopted in paper V and taken 
from Caughlan et al. (\cite{Caughlan85}). 
Finally, let us recall that as in paper V, we adopted the Schwarzschild criterion for convection 
without overshooting.

Thus, with respect to paper V, in addition to the improvements of the physics describing the effects of rotation
on the mixing of the chemical ele\-ments and on the mass loss rates, we have updated
the mass loss rates (see previous section), the initial composition, the opacities
and the nuclear reaction rates.
In order
to evaluate the importance of the changes which are
 not linked to the physics of rotation, we have computed
a non--rotating 20 M$_\odot$ model at solar metallicity with the physics included in this paper (see  Fig.~\ref{HRvingt} below). 
We obtain that the present 20 M$_\odot$ model at solar metallicity follows
a very similar track in the HR diagram as the one presented in paper V.
With respect to the model of paper V, the MS lifetime is increased by about 7\%, while the helium--burning phase
is decreased by the same amount. This results in part from the slightly increased initial hydrogen abundance
in the present model (X=0.705 while in paper V, X = 0.680). The mass loss during the MS phase has been significantly reduced
in the present paper. Indeed, the present 20 M$_\odot$ model at solar metallicity has 
lost during the MS phase a little more than
one third of a solar mass, while the mass of the
corresponding model presented in paper V decreased by about one solar mass during the same period.

\begin{figure}[tb]
  \resizebox{\hsize}{!}{\includegraphics{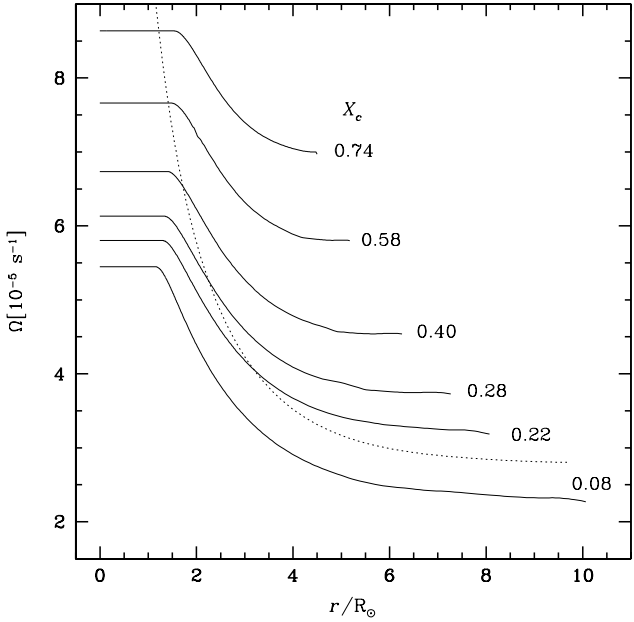}}
  \caption{Evolution of the angular velocity $\Omega$ 
 as a function of the distance to the center
in a 20 M$_\odot$ star with $v_{\rm ini}$ = 300 km s$^{-1}$ and
$Z = 0.004$. 
$X_c$ is the hydrogen mass fraction at the center.
The dotted line shows the profile when the He--core contracts at the end
of the H--burning phase.}
  \label{omegar}
\end{figure}

\begin{figure}[tb]
  \resizebox{\hsize}{!}{\includegraphics{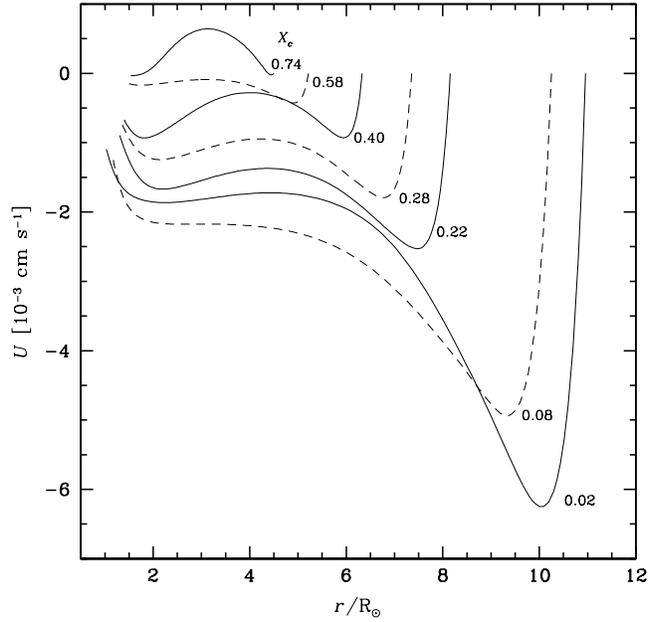}}
  \caption{Evolution of $U(r)$ the radial term of the vertical component of the
velocity of meridional circulation for the
same model as in Fig.~\ref{omegar}. $X_c$ is the hydrogen mass 
fraction at the center.}
  \label{Ur}
\end{figure}

\section{The evolution of the internal rotation
and meridional circulation}

The initial convergence of the internal $\Omega$--profile was discussed
in paper V and here the results are the same.
 The main point was that the 
$\Omega$--profile  converges very quickly in about 1 to 2\%
of the MS lifetime towards an equilibrium profile. The same fast
convergence occurs
 for the internal distribution of $U(r)$, the radial term of 
the vertical component
of the meridional circulation. Then, during the MS phase
$\Omega(r)$ decreases slowly
keeping a small degree of differential rotation, which plays an
essential role in the shear effects and the related transport mechanisms.
We recall here the interesting result shown in paper V that,
in contradiction to the classical Eddington--Sweet theory, 
the velocity $U(r)$ does not depend very much on the 
rotational velocities.

Fig.~\ref{omegar} shows the evolution of the 
$\Omega$--profile during the  evolution on the MS
of a 20 M$_{\odot}$  star with $Z$ = 0.004 and an initial 
rotation velocity $v_{\mathrm{ini}}$ = 300 km s$^{-1}$.
Let us point out some
differences with respect to the models at $Z$ = 0.020. 
For the same initial rotational velocities $v_{\mathrm{ini}}$,
at a given central H--content X$_{\mathrm{c}}$, the values of
$\Omega(r)$ are higher at $Z$ = 0.004.
One first reason is rather trivial, i.e. because the radius is smaller
at lower $Z$, the same $v_{\mathrm{ini}}$ corresponds
to a larger angular velocity $\Omega$.
Indeed, the radius of  a 20 M$_{\odot}$ star 
at $Z$ = 0.004  is smaller, being equal to 0.8
of the radius of a same mass star at $Z$ = 0.020. This ratio of 0.8
keeps closely the same during the whole MS phase.
Another reason for the higher $\Omega(r)$ lies in the smaller
mass loss rates, which  lead to smaller losses of angular
momentum, thus  favouring a higher internal rotation
at the end of the MS phase. 

The  internal gradients of $\Omega$  are higher at lower
Z due to the higher compactness of the star. This means that the 
outwards transport of chemical elements by shears
will be favoured at lower Z.  However, these larger shears
do not destroy the $\Omega$--gradient, since quite generally
 shears are
 much less important than meridional circulation for transporting the
angular momentum and
shaping $\Omega(r)$.

Fig.~\ref{Ur} shows the evolution of $U(r)$  in the same model.
 $U(r)$ is initially positive in the  interior, but
progressively  the 
fraction of the star where  $U(r)$ is negative is growing. This is
due to the Gratton--\"{O}pik term in Eq. (2),
which favours a negative $U(r)$ in the outer layers, 
when the density decreases. This negative velocity causes
 an outward transport of the angular momentum, as well as the
shears. Due  to the  higher density in the 
envelope at $Z$ = 0.004, the Gratton--\"{O}pik term is less important 
and the values of $|U(r)|$ are  smaller than at a solar composition.
However, this is partly compensated by the smaller stellar radius
which reduces the characteristic time for transport.
 In the model of 20 M$_{\odot}$ at $Z$ = 0.004, 
we have the timescales
$\tau_{\mathrm{circ}} \sim R/U = 1.5 \cdot 10^7 $ yr. and 
$\tau_{\mathrm{shear}} \sim R^2/D_{\mathrm{shear}} = 8.0  \cdot
10^7 $ yr. This confirms that  the
 circulation $U(r)$ is, as for models at $Z$ = 0.02,
  more important than 
the shears for transporting the angular momentum.
In addition, we see that 
$\tau_{\mathrm{circ}}$ is of the same order as the MS lifetime 
of $9.7 \cdot 10^6$ yr., which
allows a significant transport to occur.

\begin{figure}[tb]
  \resizebox{\hsize}{!}{\includegraphics{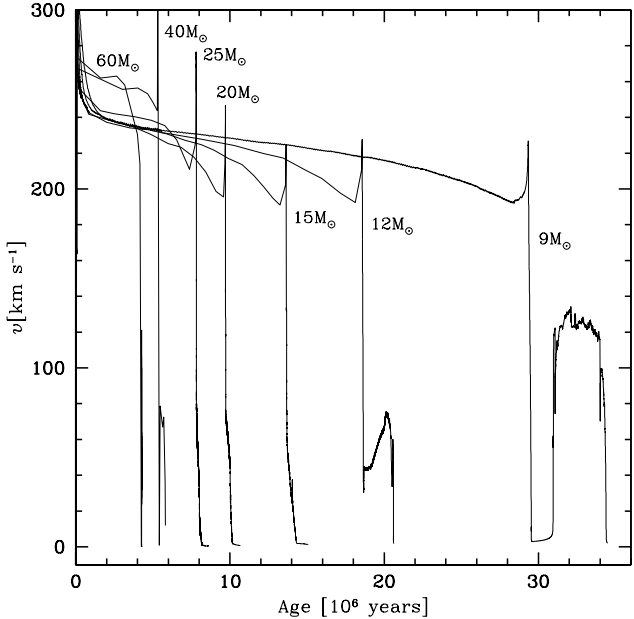}}
  \caption{Evolution of the surface equatorial velocity as 
a function of time for stars of different initial masses
with $v_{\mathrm {ini}}$ = 300 km s$^{-1}$ and $Z$ = 0.004. 
}
  \label{v/age}
\end{figure}

\section{The evolution of the surface rotation velocities}

The evolution of the rotational velocities
$v$ at the stellar surface
is shown in Fig.~\ref{v/age}. We notice that
\emph{ for all masses considered here,  $v$
remains almost constant until the end of the MS phase}.
Starting from an initial velocity $v_{\mathrm{ini}} = 300 $
km s$^{-1}$ on the ZAMS,  the initial convergence brings
$v$ to about 250 ($\pm$ 10 km s$^{-1}$) and these
 values decrease to about 200 km s$^{-1}$
for stars in the range of 9 to 25 M$_{\odot}$
at the end of the MS phase.
This behaviour is very different from that found for models 
with $Z$ = 0.02 (paper V), where  the values of $v$
for the large masses  decreased
 drastically (a value as low as 30 km s$^{-1}$ was found for the 
60 M$_{\odot}$ star at the end of the MS). The reason for the
decrease of $v$ at $Z$ = 0.02 was the large
mass loss rate $\dot{M}$, while here at $Z$ = 0.004 the 
$\dot{M}$--rates are much smaller and the stars retain
 most of their angular momentum. This behaviour
is reminiscent of that found in models 
with solid body rotation by Sackmann \& Anand (\cite{Sa70})
and by Langer (\cite{La97,La98}).

\begin{figure}[tb]
  \resizebox{\hsize}{!}{\includegraphics{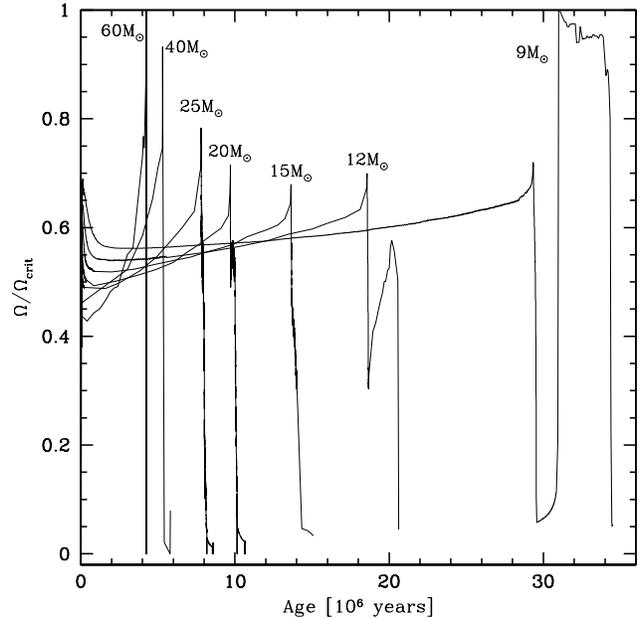}}
  \caption{Evolution of the ratio $\Omega/\Omega_{\rm crit}$ 
of the angular velocity to the break--up angular velocity
at the stellar surface for stars of different masses 
at $Z$ = 0.004.
}
  \label{omegage}
\end{figure}

Fig.~\ref{omegage}
shows the evolution of the fraction
of the critical angular velocities
$\frac{\Omega}{\Omega_{\rm{crit}}}$;
these results   are
striking and potentially  very important.
For the model of 9  M$_{\odot}$,
this fraction remains almost constant, but for the 
higher masses, the ratio $\frac{\Omega}{\Omega_{\rm{crit}}}$
grows a lot and may even 
reach 1.0 during the overall contraction phase at the end of the 
MS phase. This is  the opposite of the results of the
models with  $Z$ = 0.02,
where the fraction $\frac{\Omega}{\Omega_{\rm{crit}}}$
strongly decreased for the largest masses due to their high mass
loss, which  removed a huge amount of angular momentum.
In the present models at $Z$ = 0.004, mass loss 
is insufficient to remove enough angular momentum.

We recall that, under the hypothesis of  local conservation
of angular momentum, the ratio $\frac{\Omega}{\Omega_{\rm{crit}}}$
would decrease for a growing radius.
Thus, the growth of $\frac{\Omega}{\Omega_{\rm{crit}}}$
for the largest masses
in Fig.~\ref{omegage} results essentially from the larger outwards transport 
of angular momentum  by circulation.
Indeed, we notice that $U(r)$ in a 40  M$_{\odot}$
with $Z$ = 0.004 near the end of the MS
 reaches $-6 \cdot 10^{-2} \mathrm{cm\  s}^{-1}$, compared to 
$-0.6 \cdot 10^{-2} \mathrm{cm\  s}^{-1}$ in the 20 M$_{\odot}$
illustrated in Fig.~\ref{Ur}. The large negative velocities
in the 40 M$_{\odot}$ model at $Z$=0.004
transport more angular momentum outwards and thus
explain the rise of $\frac{\Omega}{\Omega_{\rm{crit}}}$
at the surface for large masses in Fig.~\ref{omegage}.

Now, why is $U(r)$ more negative in larger masses ?
From Eq. (1) we see that for given values of 
$(\nabla_{\mathrm{ad}}- \nabla_{\mathrm{rad}})$ and $\nabla_{\mu}$,
$U(r)$ behaves essentially like

\begin{equation}
U(r)  \sim  \frac{L}{\beta g M} 
\left[ 1 - \frac{{\Omega^2}}
{2\pi G{\overline{\rho}}}\right]  
\left( \frac{\Omega^2r^3}{GM}
\right) .
\end{equation} 

\noindent
For increasing stellar masses, we have a smaller value of the gravity
$g$, a smaller value of $\beta$ the ratio of
 $P_{\mathrm{gas}}$ to the total pressure $P$,
a larger $L/M$ and a smaller density. All these 
differences contribute
to make the values of $U(r)$ more  negative. 
Together with the smaller mass loss, these are  the physical
 reasons for the growth of
the surface ratios $\frac{\Omega}{\Omega_{\rm{crit}}}$
with the stellar masses.
These results suggest that, if the initial distributions of the 
rotational velocities 
do not have a higher  fraction of slow rotators at lower metallicities
(a point which is still  unknown),
then we may expect that \emph{at lower Z, as in the SMC, a much larger
fraction of O-- and B--type stars are close to break--up velocities}. This 
suggests that rotation is not necessarily  a secondary 
parameter for the evolution of massive stars at low $Z$.

\begin{figure}[tb]
  \resizebox{\hsize}{!}{\includegraphics{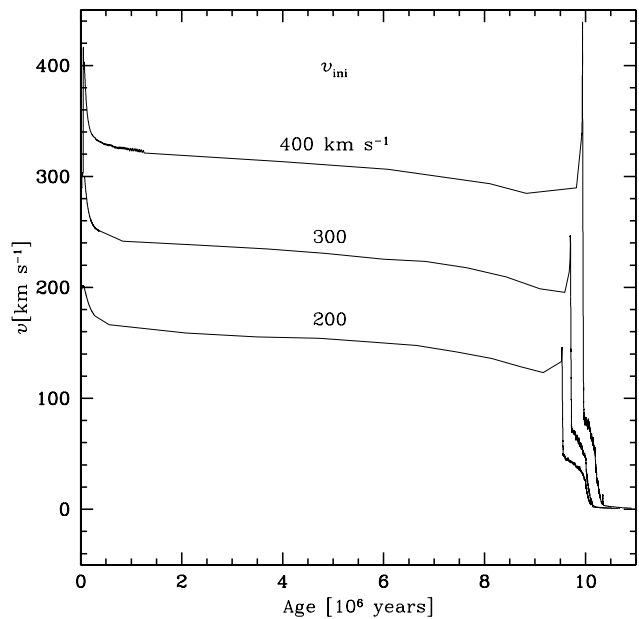}}
  \caption{Evolution of the surface equatorial velocity as 
a function of time for 20 M$_\odot$ stars at $Z$ = 0.004 with different
initial velocities. 
}
  \label{v20age}
\end{figure}

Fig.~\ref{v20age} shows the evolution of the surface rotational velocities
$v$
as a function of time for different initial velocities in the case
of the model of 20 M$_{\odot}$ at $Z$ = 0.004.
We notice the rather parallel behaviour of the evolution
of $v$. This is rather different from the 
models at $Z$ = 0.02, which  showed a certain degree
 of convergence of  $v$ at the end of the MS,
due to the large mass loss rates.

With some necessary reservations, let us now make a few
speculations.  For zero metallicity models, 
the MS phase is shifted by
0.27 dex in $T_{\mathrm{eff}}$ with respect to the models at solar
composition,  which means radii smaller by a factor of 3.4
at a given luminosity for $Z$ = 0.
For the models at $Z$ = 0.004, the blue shift amounts
 only to 0.04 dex, implying a radius smaller by a factor of 1.2.
Thus, we see that the degree of compactness at very low $Z$
is much stronger.  
Thus, the  models at $Z$ = 0 
will likely amplify the behaviour shown by the models
at $Z$ = 0.004, due to their higher compactness
and  their smaller
 mass loss rates. 
 Thus, we may wonder whether \emph{ a large fraction, 
if not most, massive stars at very low $Z$ do not reach break--up
velocities during their MS phases}. The observations
of a much higher fraction of Be--stars (cf.
Maeder et al. \cite{maegremer})
in the LMC and SMC compared to the Milky Way
tend to support this view.

Finally, we  notice in Fig. 4 the occurence of
the so--called spin--up effect  in the 9 M$_{\odot}$
on the blue loop:  as the star  is contracting,  
its rotation strongly accelerates. This effect has been described  by
Heger \& Langer (\cite{HL98}). The envelope 
accelerates its rotation, due to both its contraction and the
fact that the convective zone is receding, as shown by Heger \&
Langer. 

\section{Models with zero rotation}
 
The evolutionary tracks and the lifetimes of 
the non--rotating tracks at $Z$ = 0.004
are presented in Figs.~\ref{HRge} and \ref{HRnr} 
and in Table~\ref{tbl-1}. With respect to the 
previous Geneva grids of stellar models
at $Z$ = 0.004 (Charbonnel et al. \cite{Charbonnel};
 Meynet et al. \cite{Mey94}), 
those presented here  show the following main differences:
the MS width is reduced, typically at $\log T_{\rm eff}$ = 4.4, 
the width in $\Delta \log L$/L$_\odot$ is decreased
by 0.34 dex; for initial mass star models below 40 M$_\odot$, 
the MS lifetimes are shorter by about 11\% for the
9 M$_\odot$ model and by 1.5\% for the 40 M$_\odot$ model; the ratios of the He-- to the H--burning lifetimes are greater
by about a factor two for the 9 M$_\odot$ model and by 7\% for 
the 40 M$_\odot$ model.
All these effects are quite consistent with the fact that, 
in the present work, we did not include any overshooting
(see e.g. Bertelli et al. \cite{be85};
 Maeder \& Meynet \cite{MM89}; Chin \& Stothers \cite{Ch90};
 Langer \& Maeder \cite{LM95} for detailed
discussions of the effects of overshooting on the stellar models). 
For initial masses superior to about  60 M$_\odot$, the effects of the stellar winds become very important and dominate the effects
due to a change of the criterion for convection.
Qualitatively si\-milar differences are obtained when the present results are compared with the stellar models at $Z$ = 0.004
computed with overshooting by
Fagotto et al. (\cite{Fagotto}) and Claret \& Gimenez (\cite{Claret}).

The models with initial masses between 10 and 12.5 
M$_\odot$ are peculiar, they show a behaviour in the HR diagram
intermediate between the cases of stars presenting
 a well developed blue loop (as the 9 and the 10 M$_\odot$
models in Fig.~\ref{HRnr}) and 
the case of more massive stars
 which do not produce any blue loop,
 but begin to burn their helium in their core
at a high effective temperature while
 they cross the HR diagram for the first
 time (as the 12.5 and 13 M$_\odot$ in Fig.~\ref{HRnr}).
The models in the transition mass range ({\it i.e.} the 10.5, 11 and 12 M$_\odot$ models in Fig.~\ref{HRnr}) present a ``partial
blue loop'' which starts at a high effective temperature and then
the models  spend nearly their whole helium burning 
phase  in the blue (see Sect. 7).
Such a behaviour has also been found by 
Charbonnel et al. (\cite{Charbonnel})
 for their 15 M$_\odot$ model and by Claret \&
Gimenez (\cite{Claret}) for stars with initial masses between about 16 and 30 M$_\odot$. In this last case, the exact mass range depends on the
adopted value for the initial abundance of helium. 
These results seem to indicate that
 an extension of the H--burning core, e.g. by overshooting, shifts the transition range to higher initial masses, which is quite consistent.
The models by Fagotto et al. (\cite{Fagotto}) do not show 
such a behaviour between 15 and 30 M$_\odot$.
This may be due to the fact that 
these models were 
computed with a much greater overshooting parameter
than the two grids mentioned above.

\section{HR diagram, mass--luminosity relation and lifetimes}

\subsection{A brief review of the effects of rotation on the stellar models}

\begin{figure*}[tb]
  \resizebox{\hsize}{!}{\includegraphics[angle=-90]{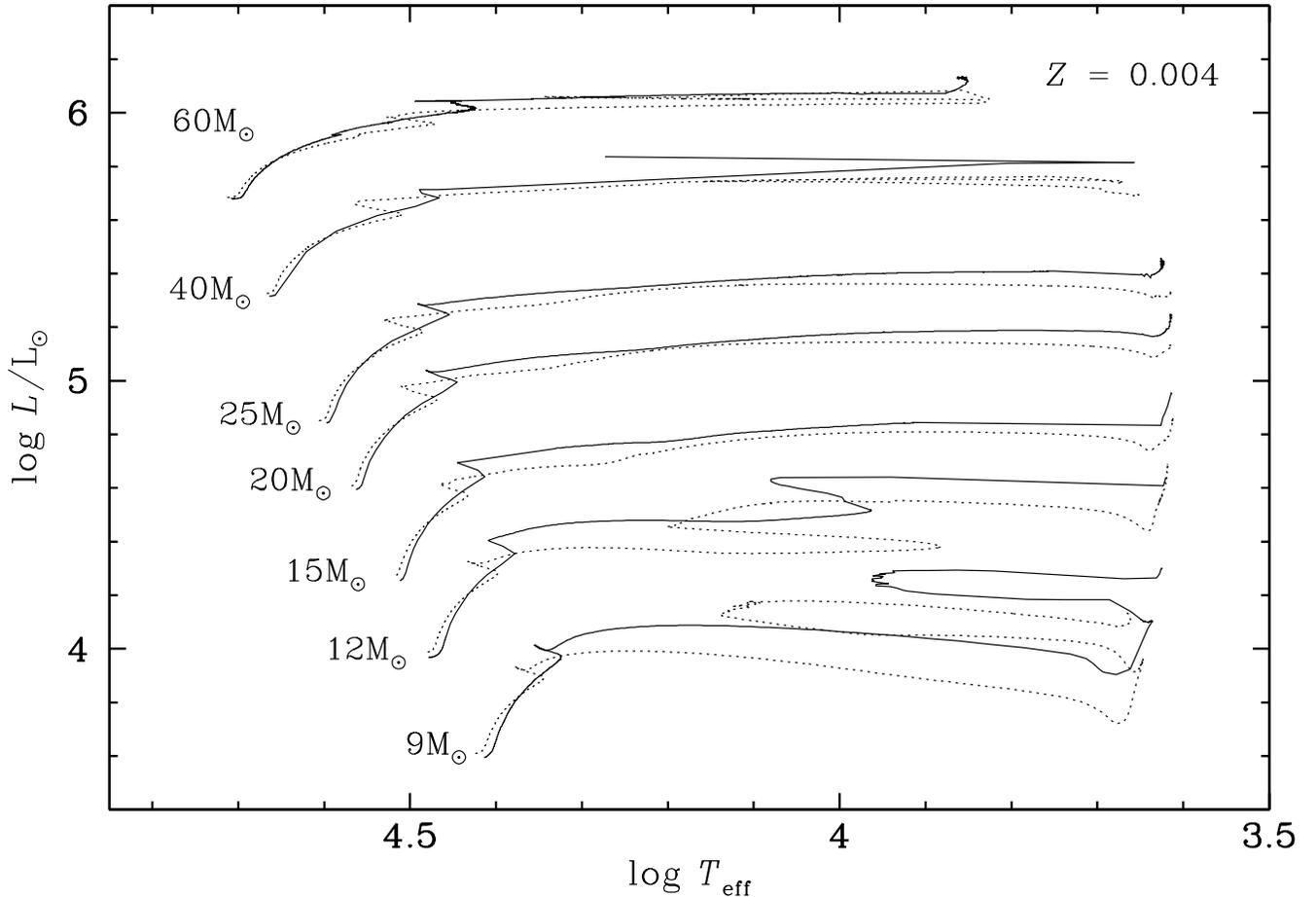}}
  \caption{Evolutionary tracks for non--rotating 
(dotted lines) and
rotating (continuous lines) models for a metallicity $Z$ = 0.004. 
The rotating models
have an initial velocity $v_{\rm ini}$ of 300 km s$^{-1}$.
For the rotating 60 M$_\odot$ model, only part of the evolution is plotted.}
  \label{HRge}
\end{figure*}

Many effects induced by rotation at solar metallicity are also found
here at lower metallicity. These effects have
 been described in detail by Heger et al. (\cite{HL00}),
Maeder \& Meynet (\cite{MMARAA}) and Meynet \& Maeder (\cite{MMV}).
 Thus we will be very brief here. 
Rotation modifies the evolutionary tracks in the HR diagram through
the following main physical effects:
rotation lowers the effective gravity $g_{\rm eff}$; 
it enlarges the convective cores and
smoothes the chemical gradients in the radiative zones (the main effect);
it enhances the mass loss rates;
it produces atmospheric distortions.
Since the star has lost its spherical symmetry, the tracks
may also change in appearance depending on the angle of view. Here 
we  suppose an average
angle of view, as was done in paper V. 

As a consequence of these effects, depending 
on the degree of mixing (see below), the 
evolutionary tracks on the MS are extended towards
lower effective temperatures 
(as would do a moderate overshooting) and/or are made 
overluminous.
The evolution towards the red supergiant stage is favoured (see Sect. 7), as well as 
the evolution into the WR phase.
Moreover, as will be discussed below, the MS lifetimes are increased and
the surface abundances are modified.

Since there are different rotational velocities, a star of given initial
mass and metallicity can follow different tracks corresponding to various
initial rotational velocities.
 To minimize the number of tracks 
to be computed, we have to choose
a value of the initial velocity which 
produces an average velocity during the MS not too far from the 
observed value. The problem here is that one does not know 
what this average rotational velocity is
for OB stars at the metallicity of 
the Small Magellanic Cloud (SMC).
 In the absence of such an information, we
adopt here the same initial rotational velocity as in
 our grid at $Z$ = 0.020, namely
the value $v_{\rm ini}$ = 300 km s$^{-1}$.
Defining a mean equatorial velocity  $\overline{v}$ during the MS as in
Meynet \& Maeder (2000), such a value for $v_{\rm ini}$
 corresponds to values of $\overline{v}$ 
between 220 and 260 km s$^{-1}$. These values are close
 to the average rotational velocities observed
for OBV type stars at solar metallicity, which are
 between 200--250 km s$^{-1}$.

\subsection{The HR diagram}

Figure~\ref{HRge} shows the evolutionary tracks of non--rotating and rotating
stellar models for initial masses between 9 and 60 M$_\odot$. One sees that 
the MS width is increased, as would result
from a moderate overshoot.
Let us recall here that two counteracting effects of rotation
affect the extension of the MS band (see paper V).
On one hand, rotational mixing
brings fresh H--fuel into the convective core, 
slowing down its decrease in mass
during the MS. This effect produces a 
more massive He--core at the end of the H--burning phase
and this favours the extension of the tracks towards lower effective temperatures. 
On the other hand, rotational mixing transports
helium and other H--burning products (essentially nitrogen)
into the radiative envelope. The He--enrichment lowers the opacity. This
contributes to the enhancement of the stellar 
luminosity and favours a blueward track.
Clearly here, the first effect dominates over the second one. 
  
In this context, it is interesting to recall that, due to the account of
horizontal turbulence
in the present models, the mixing of the chemical elements 
by the shear has been effectively 
reduced in the regions of steep $\mu$--gradient
with respect to paper V (see Sect. 2). 
This favours, for a given initial rotational 
velocity, an extension of the MS towards lower effective temperatures. 
Indeed when rotational mixing is decreased, either
as a consequence of the reduction of the initial velocity or
by a reduction of the shear diffusion coefficient 
as in the present models, the time required for 
helium mixing in the whole radiative envelope is considerably increased, 
while the time for hydrogen to migrate into the convective core,
although it is also increased, remains nevertheless
relatively small since hydrogen
just needs to diffuse 
through a small amount of mass to reach the convective core
(see Meynet \& Maeder \cite{MMV}). Thus, some increase in 
the size of the core 
results, while the effect on the helium abundance in the envelope is
not significant. The same kind of effect can be  seen in the models
by Talon et al. (\cite{Ta97}).
Of course the $\mu$--gradients are 
strong near the core and can slow down the diffusion process mentioned above, but on the other
hand, the efficiency of the diffusion of hydrogen will
 also increase with the increasing H--abundance gradient
at the border of the core. These are the various reasons why  
the numerical models show that, for a 
 moderate rotational
mixing, the effect of rotation on the convective core mass
 overcomes the effect of helium diffusion in the envelope. 

Let us recall that some core overshooting was needed
in stellar models in order to reproduce 
the observed MS width (see Maeder \& Mermilliod \cite{mm81};
 Maeder \& Meynet \cite{MM89}). Typically,
the width of the MS band in 
$\Delta \log L$/L$_\odot$ at log $T_{\rm eff}$ = 4.4 
obtained from the present non--rotating models
is 0.85, while a value of about 1.20 is required to reproduce the observation.
Such a width can be reproduced with a moderate amount of overshooting (see Schaller et al. \cite{Schall92}).

Rotation as we just saw above
produces also a  wider MS. Indeed, the width of the MS band in
 $\Delta \log L$/L$_\odot$ at log $T_{\rm eff}$ = 4.4 for the rotating models
presented in Fig.~\ref{HRge} is slightly superior to 1. 
This comparison shows that if the $v_{\rm ini}$ = 300 km s$^{-1}$
models are well representative of the average case, then rotation alone might account for about  half
of the MS widening required by the observations.
This is only a rough estimate. The MS extension is different for different initial rotational velocities (see Fig.~8),
it will also be slightly different depending on the angle of 
view of the stars. Therefore
the evaluation of the contributions 
of rotation and  overshooting to the
widening of the MS band requires the
 computations of numerous models for various initial masses and
rotational velocities so that detailed population synthesis models can be performed.
However, we can say with some confidence
that rotation decreases by about a factor of 2
the amount of overshooting needed to reproduce the observed MS width,
as was already proposed by Talon et al. (\cite{Ta97}).

One striking difference between non--rotating and rotating models after the MS
concerns the fraction of the helium burning phase spent as a red supergiant. This point
will be discussed in detail in the next section. 
Let us also mention here that
rotation shortens the blue loops of the 9 M$_\odot$ model. 
This is a consequence
of the more massive helium cores existing at the end of 
the H--burning phase in rotating models, as well as of the addition of
helium near the H--burning shell (cf. Sect. 7.2).
As in the case of the 9 M$_\odot$ model and for the same reason, rotation reduces the extension of the ``partial'' blue loop associated with
the 12 M$_\odot$ model.

For the rotational velocities considered here, 
no model enters
the WR phase during the core H--burning phase.
But after the MS phase, the rotating 40 M$_\odot$ model enters the WR phase while its non--rotating counterpart
does not. This illustrates the fact that rotation decreases
the minimum initial mass for a single star to become a WR star
 (Maeder \cite{ma87}; Fliegner \& Langer \cite{fl95};
 Maeder \& Meynet \cite{MMARAA}; Meynet \cite{me00}).
We shall not develop this point further here since it 
will be the subject of a forthcoming  paper.

\begin{figure}[tb]
  \resizebox{\hsize}{!}{\includegraphics[angle=0]{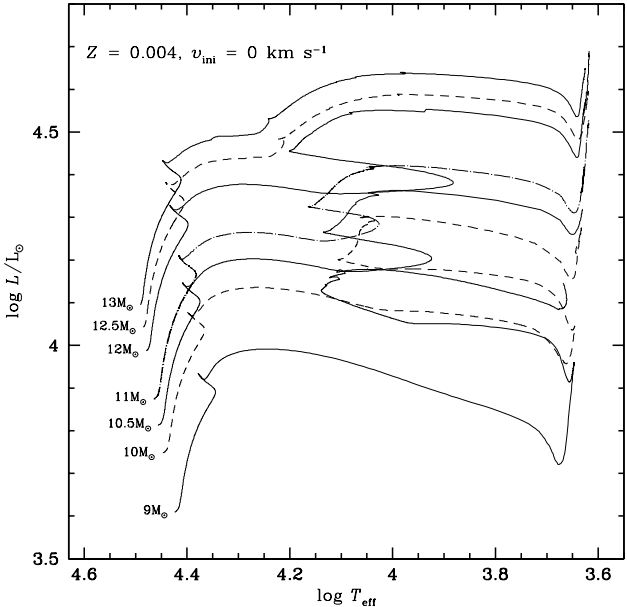}}
  \caption{Evolutionary tracks for non--rotating 
stellar models at $Z$ = 0.004 in the transition region between
stars with and without blue loops.
 The initial masses are indicated in solar
masses.
}
  \label{HRnr}
\end{figure}

\subsection{Differences between rotating tracks at $Z$ = 0.004 and 0.020}

\begin{figure}[tb]
  \resizebox{\hsize}{!}{\includegraphics{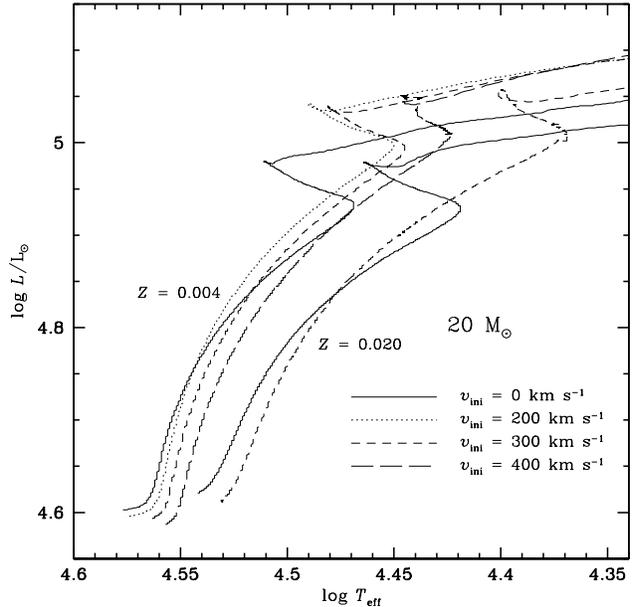}}
  \caption{Evolutionary tracks for rotating 20 M$_\odot$ models with different
initial velocities and various initial metallicities. 
The initial velocities $\upsilon_{\rm ini}$ are indicated. 
See Table 1 for more details on the models at Z = 0.004.}
  \label{HRvingt}
\end{figure}

Since the physics has been improved with respect
to paper V (see Sect. 2), we cannot directly
 compare the solar tracks of paper V with the present ones.
Therefore
we computed one 20 M$_\odot$ model at solar metallicity with and without rotation with exactly the same
physics as in the present paper. These models are plotted together with the 20 M$_\odot$ models at $Z$ = 0.004
in Fig.~\ref{HRvingt}. 

From this figure, one sees that rotation at low metallicity has
similar effects than at solar metallicity. For instance, 
the increase due to rotation of the He--core masses
 at the end of the H--burning phase is similar
at $Z$ = 0.020 and $Z$ = 0.004. Typically, for
the 20 M$_\odot$ models shown on Fig.~\ref{HRvingt}, one has
that in the non--rotating models, both at $Z$ = 0.020 and $Z$ = 0.004, the helium cores contain 
26\% of the total mass at the end of the MS phase. In the rotating models with $v_{\rm ini}$ = 300 km s$^{-1}$, 
this mass fraction is enhanced up to values between 30 -- 32 \%.

Due to the distribution of the initial velocities and of the orientations of the angles of view,
rotation induces some scatter of the luminosities
 and effective temperatures at
the end of the MS phase (see paper V).
One observes from Fig.~\ref{HRvingt} that, at low metallicity,
for a given initial velocity, the extension
 towards lower effective temperatures due to rotation 
is slightly reduced (compare the tracks for $v_{\rm ini}$ = 300 km s$^{-1}$).
Thus, at low $Z$, our models show that, 
if the initial rotation velocities are 
distributed in the same manner as at solar metallicity, 
the scatter of the effective 
temperatures and luminosities at the end of the MS
will be reduced.

\subsection{Masses and mass--luminosity relations}

When rotation increases, the actual masses at the
 end of both the MS and the He--burning phases
become smaller (cf. Tables~1).
Typically the quantity of mass lost by 
stellar winds during the MS is enhanced by
45--75\% in rotating models with $v_{\rm ini}$ = 300 km s$^{-1}$. Similar enhancements were found
at solar metallicity (see paper V). 
The increase is due mainly to the direct effect of rotation on the mass loss rates (cf. Eq. 5). The higher luminosities
reached by the rotating models and their longer MS lifetimes also
contribute somewhat to produce smaller final masses.

Rotation makes the star overluminous for their actual masses.
Typically for
$v_{\rm ini}$ = 300 km s$^{-1}$, the luminosity  vs. mass ratios at the end
of the MS are increased  by 15--22\%.
 It is interesting to mention here
that even if the rotating and non--rotating tracks
 in the $\log L$/L$_\odot$ vs. $\log T_{\rm eff}$ plane
are very similar, they may present large differences in
the $\log g_{\rm eff}$ vs. $\log T_{\rm eff}$ plane where $g_{\rm eff}$
is estimated at an average
orientation angle as in paper V.
The large difference in log $g_{\rm eff}$ 
for similar masses, luminosities and effective
temperatures
comes from the fact that 
the effective gravity of the rotating model differs from the
gravity of the non--rotating model by an amount equal to the centrifugal
acceleration. 
Typically, the two 40 M$_\odot$ tracks plotted on Fig.~\ref{HRge} which, at $\log T_{\rm eff}$ = 4.4,
differ by only $\sim$ 0.04 dex in $\log L$/L$_\odot$, 
show important differences in the $\log g_{\rm eff}$ vs. $\log T_{\rm eff}$ plane.
For instance, the rotating 40 M$_\odot$ track overlaps
the non--rotating 60 M$_\odot$ model in this plane. 
Therefore, one could expect that the attribution of a mass to an observed star
position in the $\log g_{\rm eff}$ vs. $\log T_{\rm eff}$ 
plane is very rotation dependent.
The use of non rotating tracks would overestimate the mass
 (in the example above by 50\%), and this might be a cause of
the well known problem of the mass discrepancy (see e.g.
Herrero et al. \cite{He20}). 
Let us note that in practice the effective
gravity and the  other physical quantities 
are derived from the spectral lines, which shapes and equivalent
widths are 
also affected by rotation.

\subsection{Lifetimes}

\begin{table*}
\caption{Properties of the stellar models at the end of 
the H--burning phase, at $\log T_{\mathrm{eff}} = 4.0 $
(see text) 
and at the end of the He--burning phase. The masses are in solar mass, the velocities in km s$^{-1}$, the lifetimes in million years and the helium
abundance in mass fraction. The abundance ratios 
are normalized to their initial value.} \label{tbl-1}
\begin{center}\scriptsize
\begin{tabular}{ccc|cccccc|cccc|cccccc}
\hline
    &         &         &         &         &     &      &      &      &     &      &      &      &         &         &     &      &      &       \\
    &  & & \multicolumn{6}{|c|}{End of H--burning}&\multicolumn{4}{|c|}{ $\log T_{\mathrm{eff}} = 4.0 $} &\multicolumn{6}{|c}{End of He--burning} \\
    &     &    &     &         &     &      &      &      &     &      &      &      &         &         &     &      &      &       \\    
$M$ & $v_{\rm ini}$ &  $\overline{v}$  & $t_{\mathrm{H}}$ & $M$ & $v$ & $Y_{\mathrm{s}}$ & N/C & N/O & $v$ & $Y_{\mathrm{s}}$ & N/C & N/O & $t_{\mathrm{He}}$ & $M$ & $v$ & $Y_{\mathrm{s}}$ & N/C &  N/O \\
    &   &  &         &         &     &      &      &      &     &      &      &      &         &         &     &      &      &       \\
\hline
    &    &     &        &         &     &      &      &      &     &      &      &      &        &        &     &      &      &       \\
 60 &  0 & 0   &  3.951 & 57.709  &   0 & 0.24 & 1.00 & 1.00 &   0 & 0.57 &  127 &  267 & 0.345  & 41.733 &  0  & 0.60 &  185 &  198     \\
    & 300& 259 &  4.232 & 56.415  & 307 & 0.29 & 11.3 & 9.50 &     &      &      &      &        &        &     &      &      &      \\
    &    &     &        &         &     &      &      &      &     &      &      &      &        &        &     &      &      &       \\
 40 &  0 &  0  &  4.924 & 39.066  &   0 & 0.24 & 1.00 & 1.00 &   0 & 0.24 & 1.00 & 1.00 & 0.460  & 30.222 &  0  & 0.49 &  145 & 66.0  \\
    & 300& 257 &  5.312 & 38.650  & 332 & 0.25 & 7.26 & 4.75 &  81 & 0.60 &  113 & 68.2 & 0.496  & 20.481 &  21 & 0.73 &  203 & 595     \\
    &    &     &        &         &     &      &      &      &     &      &      &      &        &        &     &      &      &       \\
 25 & 0  &  0  &  7.196 & 24.689  &   0 & 0.24 & 1.00 & 1.00 &   0 & 0.24 & 1.00 & 1.00 & 0.806  & 24.322 &  0  & 0.24 & 1.00 & 1.00  \\
    & 300& 239 &  7.809 & 24.557  & 274 & 0.24 & 4.61 & 3.25 &  56 & 0.25 & 5.42 & 3.75 & 0.752  & 20.015 & 0.6 & 0.39 & 29.5 & 18.0  \\
    &    &     &        &         &     &      &      &      &     &      &      &      &        &        &     &      &      &       \\
 20 & 0  &  0  &  8.736 & 19.833  &   0 & 0.24 & 1.00 & 1.00 &   0 & 0.24 & 1.00 & 1.00 & 1.007  & 19.690 &  0  & 0.24 & 1.65 & 1.25  \\
    & 200& 152 &  9.533 & 19.777  & 146 & 0.24 & 2.94 & 2.25 &  33 & 0.24 & 3.52 & 2.50 & 0.963  & 18.376 & 1.2 & 0.29 & 13.2 & 8.00  \\
    & 300& 229 &  9.700 & 19.750  & 244 & 0.24 & 4.94 & 3.25 &  53 & 0.24 & 5.58 & 3.50 & 0.960  & 18.039 & 0.8 & 0.31 & 17.0 & 9.50  \\
    & 400& 311 &  9.940 & 19.683  & 429 & 0.24 & 6.39 & 3.75 &  65 & 0.25 & 6.84 & 4.00 & 0.952  & 17.758 & 1.0 & 0.33 & 20.4 & 11.0  \\
    &    &     &        &         &     &      &      &      &     &      &      &      &        &        &     &      &      &       \\
 15 & 0  &  0  & 12.158 & 14.910  &   0 & 0.24 & 1.00 & 1.00 &   0 & 0.24 & 1.00 & 1.00 & 1.515  & 14.686 &  0  & 0.25 & 5.00 & 3.00  \\
    & 300& 225 & 13.641 & 14.854  & 226 & 0.24 & 6.16 & 3.50 &  40 & 0.24 & 6.68 & 3.75 & 1.389  & 14.124 & 1.2 & 0.29 & 18.0 & 8.75  \\
    &    &     &        &         &     &      &      &      &     &      &      &      &        &        &     &      &      &       \\
 12 &  0 & 0   & 16.560 & 11.969  &   0 & 0.24 & 1.00 & 1.00 &   0 & 0.24 & 1.00 & 1.00 & 2.810  & 11.902 &  0  & 0.24 & 2.84 & 1.75  \\
    & 300& 225 & 18.568 & 11.950  & 228 & 0.24 & 5.06 & 3.00 &  74 & 0.28 & 16.5 & 7.75 & 1.978  & 11.807 & 1.9 & 0.29 & 18.6 & 8.50  \\
    &    &     &        &         &     &      &      &      &     &      &      &      &        &        &     &      &      &       \\
  9 & 0  &  0  & 25.911 &  8.996  &   0 & 0.24 & 1.00 & 1.00 &   0 & 0.24 & 3.19 & 2.00 & 4.543  &  8.966 &  0  & 0.24 & 3.19 & 1.75  \\
    & 300& 222 & 29.349 &  8.993  & 225 & 0.24 & 4.23 & 2.50 & 127 & 0.25 & 9.81 & 5.00 & 4.927  &  8.542 & 2.2 & 0.25 & 10.1 & 5.25  \\
    &    &    &         &         &     &      &      &      &     &      &      &      &        &        &     &      &      &       \\
\hline
\end{tabular}
\end{center}

\end{table*}

Table~\ref{tbl-1} presents some properties of the models. Column 1 and 2 give the initial mass and the initial velocity $v_{\rm ini}$ respectively.
The mean equatorial rotational velocity $\overline{v}$ during the MS phase is indicated in column 3.
The
H--burning lifetimes $t_{\mathrm{H}}$, the masses M, the equatorial velocities $v$, the helium surface abundance $Y_{\mathrm{s}}$ and the 
surface ratios N/C and N/O at the end of the H--burning phase and normalized to their initial values are given in columns 4 to 9.
The columns 10 to 13 present some properties of the models when
$\log T_{\rm eff}=4.0$ during the crossing of the Hertzsprung--Russel 
diagram, or when the star enters into the WR phase
(for the rotating 40 M$_\odot$ models and the 
non--rotating 60 M$_\odot$ model),
or at the bluest point
on the blue loop (for the models with  M$\le 12$ M$_\odot$ ). 
The columns 14 to 19 present some characteristics of the stellar models at the end of the He--burning phase;
$t_{\mathrm{He}}$ is the He--burning lifetime.

From Table~\ref{tbl-1} one sees that for $Z=0.004$ the MS lifetimes are 
increased by about 7--13\% for the mass range between 
9 and 60 M$_\odot$ when   $v_{\rm ini}$
 increases from 0 to $\sim$300 km s$^{-1}$. 
In general, the corresponding changes in the He--burning 
lifetimes are inferior to 10\%. 
As was the case at solar metallicity, the ratios
 $t_{\mathrm{He}}/t_{\mathrm{H}}$ of the
He-- to H--burning lifetimes are only slightly decreased by rotation and
remain around 9--17\%.
At solar metallicity, the changes of the lifetimes due to rotation are quite similar.
The rotating 20 M$_\odot$ model with $v_{\rm ini}$ = 300 km s$^{-1}$, at solar metallicity, has a MS lifetime
increased by 14\% with respect to the non--rotating model. At $Z$ = 0.004, the corresponding increase
amounts to 11\%, which is not significantly different.

\section{The ratio of blue to red supergiants in the SMC}

\subsection{Model results}

The observed ratio $B/R$  of blue to red supergiants
 in the SMC cluster NGC 330
lies between 0.5 and 0.8, according to the various sources
discussed in Langer \& Maeder (\cite{LM95}). Not many new results have 
been obtained since then. New IR searches have revealed  some AGB
stars in the SMC (Zilstra et al. \cite{Zi96}) and ISO observations
(Kucinskas et al. \cite{Kuc00})
have led to the detection of an IR source in NGC 330, which may be 
a Be supergiant or a post AGB--star, but this does not change
the statistics significantly. Notice that the definition
of $B/R$ is not always the same, e.g. for 
 Humphreys \& McElroy (\cite{Hum84}),
B means O, B and A--supergiants. Here, we strictly count in the $B/R$
ratio the B star models from the end of the MS to
type B9.5 I,  which corresponds
to $\log T_{\mathrm{eff}} = 3.99$
according to the calibration by Flower (\cite{Flow96}).
 We count  as red supergiants all star models  below  
 $\log T_{\mathrm{eff}} = 3.70$ since red supergiants in the SMC
are not as red as in the Galaxy (Humphreys \cite{Hum79}). We
note that the exact definition of this limit has no influence
on the observed or theoretical $B/R$ ratios, since the evolution
through types F,G,K is always very fast.

\begin{figure}[tb]
  \resizebox{\hsize}{!}{\includegraphics{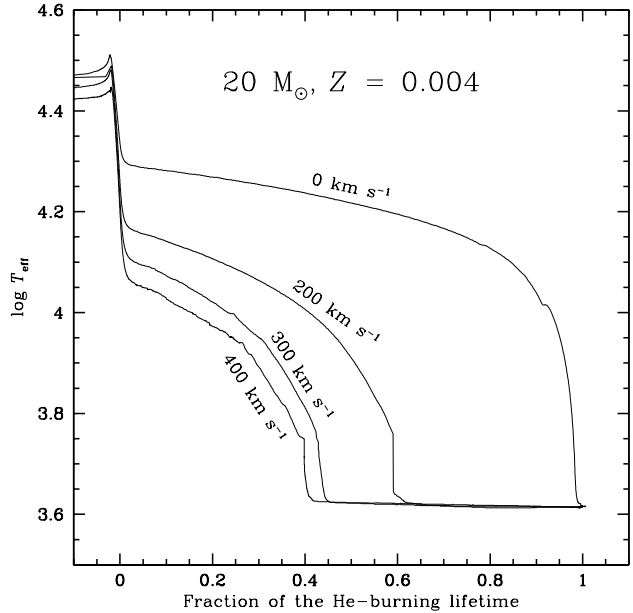}}
  \caption{Evolution of the $T_{\mathrm{eff}}$
as a function of the fraction of the lifetime spent
in the He--burning phase for 20 M$_\odot$ stars with different
initial velocities. 
}
  \label{vvv20/age}
\end{figure}

\begin{figure}[tb]
  \resizebox{\hsize}{!}{\includegraphics{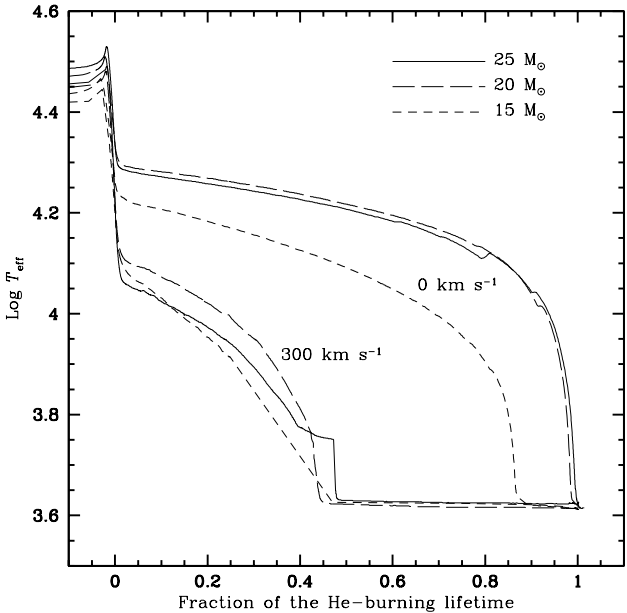}}
  \caption{Evolution of the $T_{\mathrm{eff}}$
as a function of the fraction of the lifetime spent
in the He--burning phase for 15, 20 and 25
 M$_\odot$ stars at $Z$ = 0.004 with $v_{\mathrm{ini}}$ = 0
and 300 km s$^{-1}$.
}
  \label{v152025/age}
\end{figure}

As noted by Langer \& Maeder  (\cite{LM95}), the current models 
(without rotation)
with Schwarzschild's criterion predict no red supergiants 
in the SMC (cf. Schaller et al. \cite{Schall92}). This is
also seen in Fig.~\ref{vvv20/age} which illustrates for models
of 20 M$_{\odot}$ at $Z$ = 0.004  the variations of the
$T_{\mathrm{eff}}$ as a function of the fractional lifetime in the 
He--burning phase for different  rotation velocities.
For zero rotation,  we see that the star
only moves to the red supergiants
 at the very end of the He--burning phase,
so that the $B/R$ ratio, with  the definitions given above,
is  $B/R \simeq 47$.  For  average rotational velocities 
$\overline{v}$ during the MS,  $\overline{v}$ = 152, 229 and 311
km s$^{-1}$, one has respectively  $B/R =$ 1.11, 0.43 and 0.28. 
Thus, \emph{ the $B/R$ ratios
are much smaller for higher initial rotation velocities}, as
rotation favours the formation of
red supergiants and reduce the lifetime in the blue.
We notice in particular
 that for 
 $\overline{v} =   200$ km s$^{-1}$, we have a B/R ratio  of about 0.6
well corresponding to the range of the observed values.

The $B/R$ ratios change with the stellar masses.
Fig.~\ref{v152025/age} shows for 
the models of  15, 20 and 25 M$_{\odot}$
 the changes of $T_{\mathrm{eff}}$
as a function of the fractional lifetimes in the He--burning
phase for different  rotation. For all masses, we notice that
 the non--rotating stars spend nearly the whole of 
their He--phase  as blue supergiants  and almost none as 
red supergiants. For $v_{\mathrm{ini}}$ = 300 km s$^{-1}$ (which corresponds
to about $\overline{v}$ = 220 km s$^{-1}$), 
we notice a drastic
decrease of the blue phase and a corresponding large 
increase of the red supergiant phase.

\begin{figure}[tb]
  \resizebox{\hsize}{!}{\includegraphics{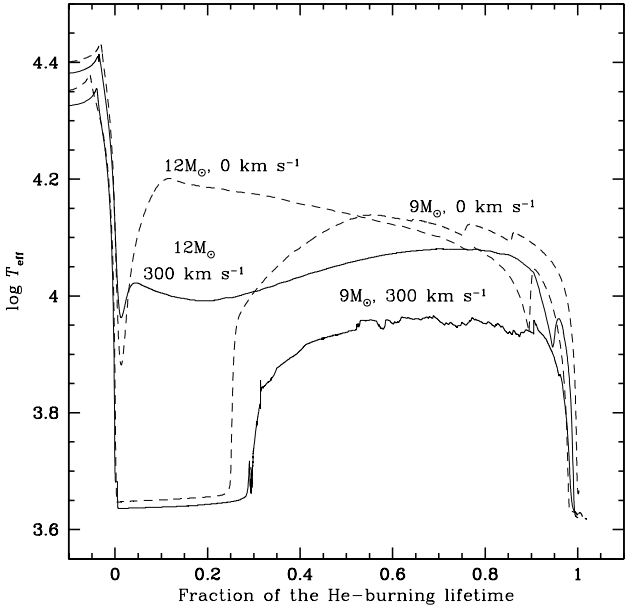}}
  \caption{Evolution of the $T_{\mathrm{eff}}$
as a function of the fraction of the lifetime spent
in the He--burning phase for 9 and 12
 M$_\odot$ stars at $Z$ = 0.004 with $v_{\mathrm{ini}}$ = 0
and 300 km s$^{-1}$.
}
  \label{v912/age}
\end{figure}

Fig.~\ref{v912/age} shows the same as Fig.~\ref{vvv20/age}
 but for the models of 9 and 
12 M$_{\odot}$. These models mark the transition from the behaviour 
of massive stars, which move  at various paces from blue to red,
to the intermediate mass stars, which go directly to the red
giant branch and then describe blue loops in the HR diagram.
At zero rotation, the 15 M$_{\odot}$ model has the ``massive
star'' behaviour and the 9 M$_{\odot}$ model shows a most 
pronounced ``blue loop''. For $v_{\mathrm{ini}}$ = 0 km s$^{-1}$, 
the 12  M$_{\odot}$
model is just in the transition between the  behaviours
of nearby models of 9 and 15 M$_{\odot}$.
The rotating model at 15 M$_{\odot}$ is first blue and then
goes to the red, while the rotating  9 M$_{\odot}$ model
 goes first to the red, then back to the blue and red again.
The  behaviour of the rotating 12 M$_{\odot}$ is also intermediate between 
these  two, with the consequence that it 
always stays more or less in the blue,
which is surprising at first sight, but well consistent
with the mentioned intermediate behaviour. 
As seen in Sect. 5, this transition zone with almost 
entirely blue models extends from about 10.5 to 12.2 M$_{\odot}$.

The rotating 9  M$_{\odot}$ model has a blue loop
smaller than for zero rotation; 
it only extends to
 the A--type  rather than to the B--type range. At a given 
$L$ and $T_{\mathrm{eff}}$, the average density 
in a rotating  model is much smaller
than in the non--rotating one, 
so that the period will be longer. The result is such that \emph{ the
application of the standard period--luminosity relation
will lead for a given observed
period to a too high luminosity, if the star was
a fast rotator on the MS}. A
more complete study of the
effects of rotation on Cepheids will be made in a further work.

Table 2 shows the $B/R$ ratios for the various relevant masses
for the models with zero--rotation and $v_{\mathrm{ini}}$ = 300 km s$^{-1}$.
Apart from the transition model of 12  M$_{\odot}$, which 
stays almost entirely in the blue as discussed above, we 
notice that  the $B/R$ ratios decrease very much with rotation,
being in the range 0.1 to 0.4 for $v_{\mathrm{ini}}=
300$ km s$^{-1}$. As noted for the 20 M$_{\odot}$ model, an
average velocity of about 200 km s$^{-1}$ corresponds to
a $B/R$ ratio of  0.6.  The order of magnitude obtained
is satisfactory, however, future comparisons in clusters will
need detailed convolution over the IMF and  the
distribution of rotational velocities in clusters
studied at various metallicities.
 This is beyond the scope of this paper and 
we now examine the effects in the internal physics
which determine the $B/R$ ratio.

\subsection{Stellar physics and the $B/R$ ratio}

There are several studies on the blue--red motions of 
 stars in the HR diagram, for example by
Lauterborn et al. (\cite{Lau71}), Stothers and Chin
(\cite{Stoth79}),  Maeder (\cite{M81}), Maeder \& Meynet (\cite
{MM89}) and recently by Sugimoto and Fujimoto (\cite{Sugi00}).
Sugimoto and Fujimoto identify several parameters
at the base of the envelope
$W, \Theta, V$ and $N$, which play a role in the redwards
evolution. Apart from $N$, which is the polytropic index,
we may note that the other parameters are all some 
function of the local gravitational potential.
$V$ is the ratio of the gravitational potential
to the thermal energy as in Schwarzschild's textbook
(\cite{Schw58}). The parameter $W$ is given by
 $W= V/U$, where $U$ is the ratio of the 
local density to the average internal density.
The parameter $\Theta$ is given by
 $\Theta = \ln (\frac{P}{\rho})_{\mathrm{c}}
- \ln (\frac{P}{\rho})_{\mathrm{env}}$, where the index 
``c'' refers to the center and ``env'' to the base of the envelope.
We can easily check that $\Theta$
is also related to the potential at the center and at the base
of the envelope, as well as to the local polytropic index.

\begin{table}
\caption{Values of the $B/R$ ratios for models with zero
rotation and for models with $v_{\mathrm{ini}}$ = 300 km s$^{-1}$.
B means strictly B--type supergiants and R means K--and
M--type supergiants (see text).
} \label{tbl1}
\begin{center}\scriptsize
\begin{tabular}{cccc}
$M_{\rm ini}$   & $ B/R $ &  $B/R$ \\

     &  $v_{\mathrm{ini}}=0$ & $v_{\mathrm{ini}}=300$\\
\hline
     &         &       &  \\
  25 &  63     &  0.30 &  \\
  20 &  47     &  0.43 &  \\
  15 &   5.0   &  0.24 &  \\
     &         &       &  \\
  12 &  20.6   & 85    &  \\
     &         &       &  \\
   9 &   2.7   &  0.10 &  \\
     &         &       &  \\

\hline
\end{tabular}
\end{center}
\end{table}

We may thus wonder whether most of the effects determining 
blue vs. red motions in the HR diagram cannot be understood,
at least qualitatively, in terms of mainly the gravitational
potential of the core. It is very desirable
to try to establish some relatively simple scheme for
understanding the results of numerical computations.
The role of the core gravitational potential 
for the inflation or deflation of the stellar radius
has been emphasized by Lauterborn et al. (\cite{Lau71})
in the case of the occurrence of blue--loops
for intermediate mass stars (see also 
  Maeder \& Meynet \cite{MM89}). We shall examine here
whether we may extend the very useful ``rules'' derived by 
Lauterborn et al. (\cite{Lau71}) to the case
of massive stars in rotation as  
studied here. We call $\Phi_{\mathrm{c}}$
the potential of the He--core, which due to a mass--radius 
relation for the core behaves as $\Phi_{\mathrm{c}} 
\sim M_{\mathrm{c}}^{0.4}$, where $M_{\mathrm{c}}$
is the core mass.
The blue--red motions in the HR diagram
mainly depend on the comparison of $\Phi_{\mathrm{c}}$
with some critical potential $\Phi_{\mathrm{crit}}(M)$,
which grows with the stellar mass.
One has

\begin{equation} 
\Phi_{\mathrm{c}}  > \Phi_{\mathrm{crit}}(M) \quad \quad \quad
\mathrm{Hayashi \quad line}
\end{equation}

\begin{equation} 
\Phi_{\mathrm{c}}  < \Phi_{\mathrm{crit}}(M) \quad \quad \quad
\mathrm{blue \quad location}
\end{equation}

\noindent
Eqs. (8) and (9) essentially apply to a steep hydrogen--profile
around the core. If this profile is mild, one should account
for additional terms (Lauterborn et al. \cite{Lau71}).
  The main effect can be
represented by a parameter $h$ increasing with
the amount $\Delta M_{\mathrm{He}}$ of helium in the 
transition region above the core
(formally, a simplified expression for  $h$ 
in the cases considered by Lauterborn et al. (\cite{Lau71})
is given by $\log h = 
8.5 X_{\mathrm{d}} M_ {\mathrm{d}}/ M$ , where 
$X_{\mathrm{d}}$ is the amplitude of the change of the
H--mass fraction over the smooth transition zone,
$M_ {\mathrm{d}}$ is the width in mass of the transition zone and
$M$ the total stellar mass). 
The value of $h$ also 
 depends on the distribution of this amount of 
helium  $\Delta M_{\mathrm{He}}$, $h$ is larger if 
the amount of helium is close 
to the shell H--burning source. However, if a sufficient
amount of helium
is brought far from the shell source, 
in the outer envelope, $h$ may even decrease as suggested 
by Lauterborn et al. (\cite{Lau71}).
This also has some consequences, as discussed below.
 With these remarks,
the two above equations become for mild H--profiles

\begin{equation} 
 h \; \Phi_{\mathrm{c}}  > \Phi_{\mathrm{crit}}(M) \quad \quad \quad
\mathrm{Hayashi \quad line}
\end{equation}

\begin{equation} 
h \; \Phi_{\mathrm{c}}  < \Phi_{\mathrm{crit}}(M) \quad \quad \quad
\mathrm{blue \quad location}
\end{equation}

\begin{figure}[tb]
  \resizebox{\hsize}{!}{\includegraphics{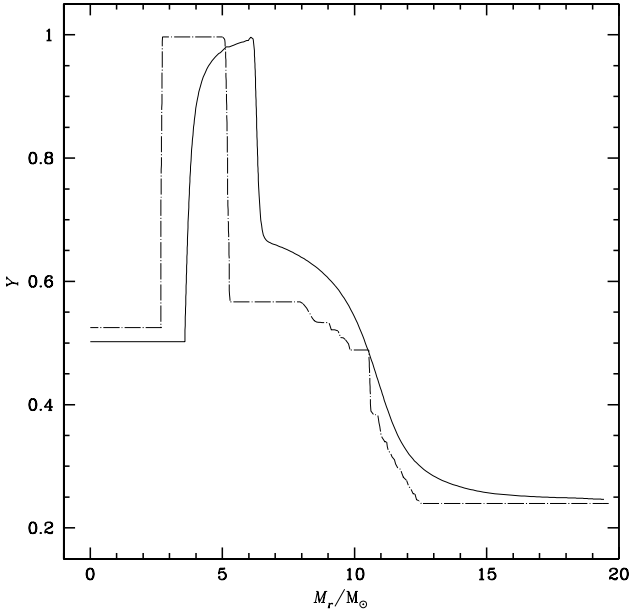}}
  \caption{ Comparison of the internal distribution of helium
in two models of 20 M$_\odot$ at the middle of the
He--burning phase. The dashed--dot line concerns the models
with zero rotation and the continuous line represents
the case with $v_{\mathrm{ini}}$ = 300 km s$^{-1}$.
}
  \label{Hedistr}
\end{figure}

 We shall now try to see whether 
we may describe correctly with relations (10) and (11)
the different physical effects influencing
the blue--red motions of massive stars:

\emph {Mass loss:} Mass loss decreases the total stellar mass
and thus $\Phi_{\mathrm{crit}}(M)$, which favours 
a motion towards the Hayashi line. $\Phi_{\mathrm{c}}$ is
not very much changed, since the size of the final He--core is not 
very different. However, there is more helium near the H--burning
shell, which increases the parameter $h$ and also favours 
the formation of red supergiants.

This description is fully consistent with the well known
fact that, due to mass
loss, the intermediate convective zone is much less important
(cf. Stothers \& Chin \cite{Stoth79}; Maeder \cite{M81}).
A convective zone imposes a polytropic index $N \simeq 1.5$,
which implies only a weak density gradient, making the stellar
radius smaller and thus  keeping the star in the blue. Thus, 
the physical connexion we have 
with the interpretation in terms of $\Phi_{\mathrm{c}}$ 
 is the following one. The larger He--burning core with respect
to the actual stellar mass together with  the higher He--content
in the H--shell region (higher $h$)
lead to a less efficient H--burning shell,
thus there is no large  intermediate convective zone and this absence
permits  a red location of the star in the HR diagram.

\begin{figure}[tb]
  \resizebox{\hsize}{!}{\includegraphics{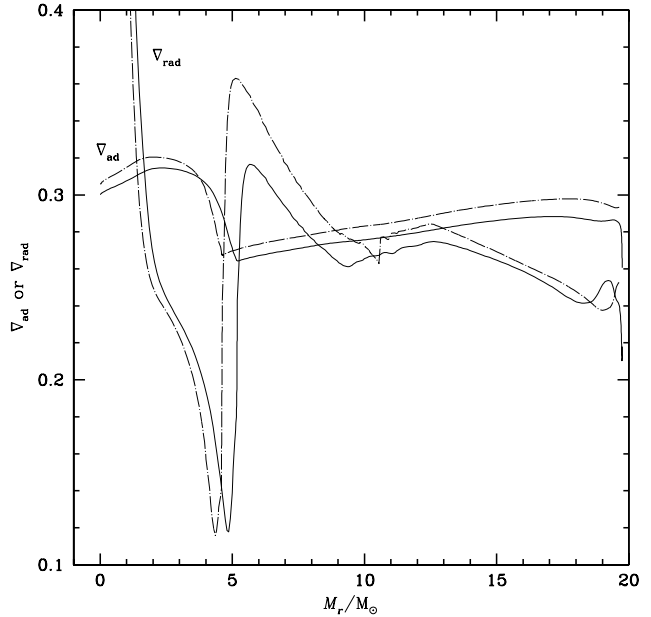}}
  \caption{ Comparison of the internal 
values of $\nabla_{\mathrm{ad}}$ and of $\nabla_{\mathrm{rad}}$ 
in models of 20 M$_\odot$ at the very beginning of the 
He--burning phase ($Y_{\mathrm{c}}$ = 0.993). 
The dashed--dot lines concern the models
with zero rotation and the continuous lines represent
the case with $v_{\mathrm{ini}}$ = 300 km s$^{-1}$.
}
  \label{Nabla1}
\end{figure}

\begin{figure}[tb]
  \resizebox{\hsize}{!}{\includegraphics{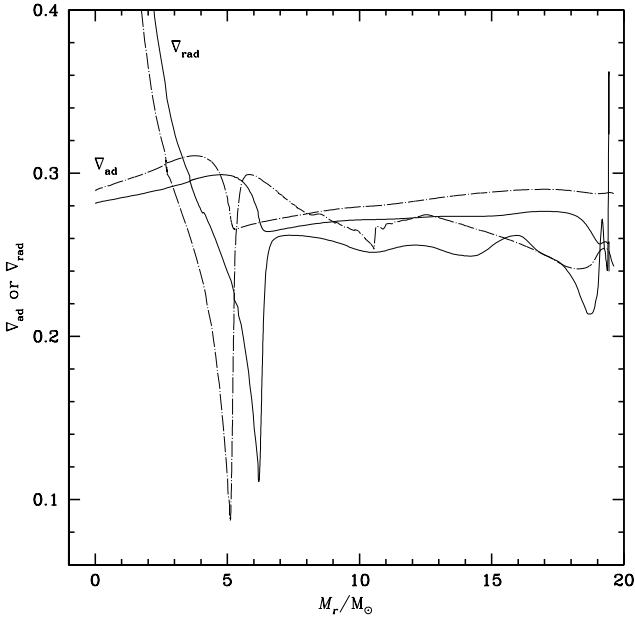}}
  \caption{ Comparison of the internal 
values of $\nabla_{\mathrm{ad}}$ and of $\nabla_{\mathrm{rad}}$ 
in models of 20 M$_\odot$ at the middle of  the 
He--burning phase (same model as in Fig. ~\ref{Hedistr}). 
Same remarks as in previous Figs.
}
  \label{Nabla2}
\end{figure}

\emph {Overshooting:} The overshooting does not change 
 $\Phi_{\mathrm{crit}}(M)$, but  $\Phi_{\mathrm{c}}$
is  increased, with no major change of the 
H--profile and thus of $h$. Clearly, overshooting is
thus favouring a redwards motion to the Hayashi line,
with the formation of red supergiants. As for mass loss,
the larger core contributes to reduce the intermediate
convective zone, which leads to the formation of red supergiants.

\emph {Lower metallicity Z:}  A lower $Z$ decreases
the mean molecular weight (essentially because the 
He--content is also lower, see Sect. 2.3). This decreases
the internal temperature and the luminosity during the
MS phase, leading to slightly smaller convective cores,
as shown by numerical models (cf. Meynet
et al. \cite{Mey94}).
This produces smaller $\Phi_{\mathrm{c}}$ which
favours a blue location, as is observed.

We also note that a lower Z means a slightly higher electron
scattering opacity (due to the higher H--content),
 which would  favour larger cores, however this   
 effect appears as a minor one in the models.

\emph {Rotation:} The effects of rotation are numerous
and subtle, and the balance  between them is 
delicate. We notice the following effects:

-- 1. A first simple effect is that rotation 
enhances the mass loss rates as described by Eq. (5)
 and this contributes to favour the formation of 
 red supergiants. However, this effect although significant
does not seem to be the dominant one.

-- 2. The mixing in the MS phase leads  to a
slight  extension of the core, which  
favours a redwards motion during the He--burning phase.
 This is just like overshooting, thus it may be very
difficult, except perhaps by asteroseismological studies of stars
with different rotations, to distinguish
between the core extension by overshooting or by rotation.
An example of this effect can be seen in Fig.~\ref{Hedistr},
which shows models of a 20 M$_\odot$ star in the middle
of the He--burning phase.
The core in the non rotating case is 2.7 M$_\odot$, while
it is 3.6 M$_\odot$ in the rotating case, i.e. 1/3 larger
in mass. Although this increase of the core is
very significant (in particular for nucleosynthesis),
its effect on the blue--red motions appears less important
than the  effect we now discuss.

-- 3. A mild mixing just  outside the core as produced 
by rotation during the MS phase clearly increases 
the amount of helium near and above the H--shell. 
This is well seen in Fig.~\ref{Hedistr}, (in the two models
shown, the H--shell
is just on the right side of the big He--peak). According to the
definition of $h$ given above, this  results in a larger
value of $h$, which leads Eq. (10) to be satisfied
and  this favours a red location in the HR diagram. 
The analysis of the sequence of the  models before
that  illustrated in Fig.~\ref{Hedistr}
shows that the different H--profile of the rotating star 
is essentially the
consequence of the  rotational mixing during the MS phase.
Thus, the rotational mixing during the MS is 
the key  effect for the formation of red 
supergiants at low $Z$.

It is satisfactory  to see that rules expressed by
 Eqs. (8) to (11) are fully consistent with
what  we can deduce by studying the consequences of a change 
of mean molecular weight. The larger He--core in the rotating 
models means that a larger fraction of the total luminosity
is made in the core (0.42 instead of 0.31 for the models
of Fig.~\ref{Hedistr}). This means that the H--burning shell
in the rotating model produces a smaller fraction of the total
luminosity and this  contributes to reduce the importance
of the convective zone  above the H--burning shell. Simultaneously,
the  higher He--content near the shell in the
rotating case leads to a
decrease in  the opacity,  and this also contributes to
reduce
the importance of the  convective zone 
associated with  the H--burning shell.
Figs.~\ref{Nabla1} and \ref{Nabla2} show
the internal values of $\nabla_{\mathrm{ad}}$ and  
$\nabla_{\mathrm{rad}}$ in 20 M$_\odot$  stars at the beginning of 
the He--burning phase  and at  the middle of this phase
respectively.
We notice the smaller convective zone associated with the
H--burning shell in the rotating model  in  Fig.~\ref{Nabla1}
and its earlier disappearance in Fig.~\ref{Nabla2}. As
long as it exists, the convective zone
 maintains  a small polytropic index over the concerned
region,  and the larger this region is, the smaller the radius.
 Only when the intermediate convective zone  disappears,  the 
star reaches  the red supergiant location in the HR diagram.
We see that the earlier disappearance of this intermediate
convective zone  in rotating stars favours their earlier
evolution towards red supergiants, consistent with
the result of  Eqs. (10) and (11).

-- 4. We may wonder what are the respective effects of mixing
on the MS and during the He--burning phase. What we have 
shown above  is just the consequence of the mixing during the MS
phase. Some tests have been made and show the following.
 If we  arbitrarily  stop the
mixing during the He--burning phase,
this makes little difference. However,   if
on the contrary  we arbitrarily
enhance the mixing during
 the He--burning phase, this   maintains
 the star in a blue location in the HR diagram. 
This result is in agreement  with the effect mentioned above
and discussed by Lauterborn et al. (\cite{Lau71}). 
The parameter $h$ normally grows with $\Delta M_{\mathrm{He}}$,
however it may reach a maximum and then decrease, 
if helium is brought far enough from the H--shell. Thus,
strong mixing in the He--burning phase by spreading 
helium throughout the star may make $h$ decrease,
 which  according to Eq. (11) leads 
to a blue location in the HR diagram.

Why does this occur physically ?  Mixing of helium also
implies mixing of hydrogen. For moderate mixing, the H--burning
shell becomes more active, due to the increase of its H--content
and $T$. This activates the intermediate convective zone associated
with the H--shell. Additional He at the surface also decreases
the opacity and favours bluer stars.
For more extreme cases of  mixing both in the MS
and subsequent phases, one  would move 
to the case of almost homogeneous stars, which 
 occurs for fast 
rotators such as the 60 M$_{\odot}$ model at $Z$ = 0.02 with a
$v_{\rm ini}$ of $\sim$ 400 km s$^{-1}$
(cf. Meynet \& Maeder
\cite{MMV}). In this case,   a bluewards evolution in the 
HR diagram occurs during the MS phase, as shown by
Schwarzschild (\cite{Schw58}).

\section{Evolution of the chemical abundances at the surface}

\begin{figure}[tb]
  \resizebox{\hsize}{!}{\includegraphics{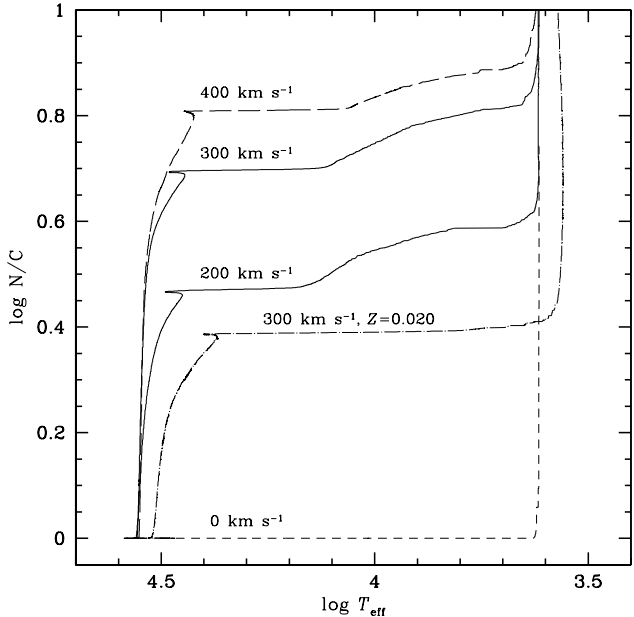}}
  \caption{Evolution as a function of $\log T_{\rm eff}$ of the abundance
ratio N/C where N and C
are the surface abundances of nitrogen and carbon
respectively. The abundance ratios are normalized to their initial value. The tracks
are for 20 M$_\odot$ at $Z$ = 0.004 with different initial rotational velocities.
The dot--long dashed line
shows the evolution of a rotating 20 M$_\odot$ at solar metallicity ($v_{\rm ini}$ = 300 km s$^{-1}$).
}
  \label{fnc20}
\end{figure}

\begin{figure}[tb]
  \resizebox{\hsize}{!}{\includegraphics{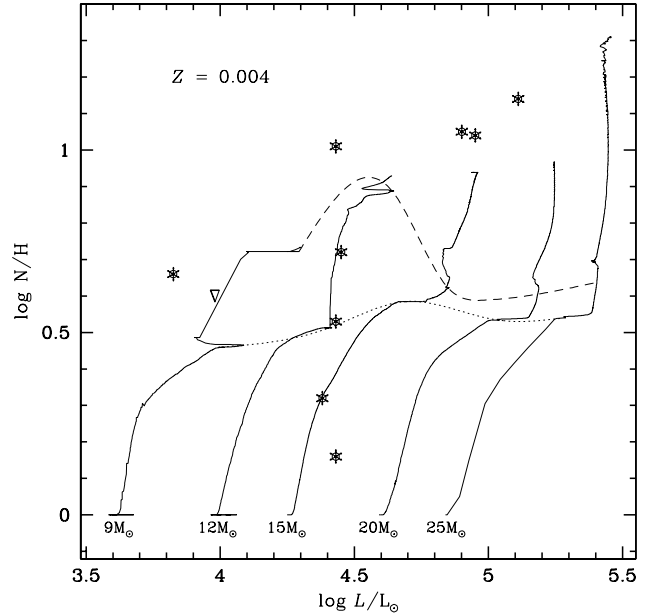}}
  \caption{Surface nitrogen to hydrogen ratios as a function of luminosity. 
The N/H ratios are normalized to their initial value.
The continuous lines represent evolutionary tracks for rotating models with $v_{\rm ini}$ = 300 km s$^{-1}$. 
The initial masses are indicated.
The dotted line indicates the N/H ratios obtained at the end of the MS
phase. The dashed line shows this same relation when 
$\log T_{\rm eff}$
is equal to 3.9. For the 9 M$_\odot$ model which presents a blue loop,
the value corresponds to the last crossing  at $\log T_{\rm eff}$ = 3.9
before the star becomes a red supergiant.
The points represent the observed values from Venn (1999), an
inverted triangle 
is plotted when only an upper limit is given. 
}
  \label{fnh}
\end{figure}

As already emphasized in previous works
(see  Maeder \& Meynet \cite{MMARAA} and references therein;
 Meynet \& Maeder \cite{MMV}; Heger \& Langer \cite{HL00}),
rotation significantly modifies the evolution of the surface
abundances. 
Fig.~\ref{fnc20} illustrates the changes 
of the nitrogen to carbon ratios N/C from the 
ZAMS to the red supergiant stage for some 20 M$_{\odot}$ stellar models.
For non--rotating stars, the surface enrichment in nitrogen
only occurs when the star reaches the red supergiant phase;
there, CNO elements are dredged--up by deep convection. 

For rotating stars, N--excesses already occur during
the MS phase. These N--excesses are not necessarily accompanied
by surface He--enrichments.
From Table~1, one sees that for the rotating models,
well before any change of the surface He--abundance occurs, the N/C and N/O ratios undergo significant changes.
This results from the effects of transport on
 the very strong gradients of the CNO elements 
which rapidly build up inside the star, (we note that mass loss at very
high  rates would do the same). 
Thus the existence of N--rich stars with a normal
 He abundance is easily explained. 

After the MS phase, the N/C ratios
 at the surface of rotating stars continue to increase
during the red supergiant phase. At low metallicity, some 
increase also occurs when the star
is crossing the HR diagram (see below). The N/C ratios obtained at the end
of the He--burning phase are much higher in rotating models than in non--rotating ones.

From Fig.~\ref{fnc20}, one sees that at a given metallicity, the
higher the initial rotational velocity, the
 more important are the surface enrichments at the end of the MS phase. 
This results from the
stronger angular velocity gradients and transport 
in faster rotating stars. 
The increase from $v_{\rm ini}$ = 200 to 300 km s$^{-1}$
produces greater relative changes than the 
increase from 300 to 400 km s$^{-1}$ (see Fig.~\ref{fnc20}).
Likely, this saturation effect results from the compensating action of two effects. On the one hand,
when the rotational velocity increases, the transport 
of the chemical species by the shear
becomes more efficient. On the other hand, when evolution proceeds, this more efficient mixing
produces smoother chemical and angular momentum gradients and thus results
in a slowing down of the transport processes. 
From Table~1, one sees that, in general, the N surface enrichments at the end
of the MS phase are higher for higher initial masses. As
explained by Maeder (\cite{ma98}) this results from the fact that, when the initial mass increases,
the ratio of the mixing timescale for 
the chemical elements to the MS lifetime decreases.

\subsection{Surface enrichments at $Z$ =0.004 and 0.020}

Very interestingly, for a given value of the initial velocity, the
lower the initial
metallicity, the  greater  the surface N--enrichments.
 The mixing of the chemical elements
is more efficient at low metallicity 
essentially because the metal poor stars are more compact and have greater angular velocity gradients.
 Thus the mixing timescales for the
chemical elements, which is about $D_{\rm shear}/R^2$, are smaller
 at lower metallicities. 

As already briefly mentioned above, the models at low metallicity present enhancements of their N/C ratios
during their crossing of the HR diagram, 
while the solar models do not present any
enrichment during this phase. This is a consequence of the fact that at low metallicity, the stars begin
to burn  helium in their core in the blue part of the HR diagram. 
This  slows down the
evolution towards the red and thus gives 
more time for  the transport processes to bring
CNO processed material at the surface, 
while the star is still in the blue.

\subsection{Comparisons with the observations}

Fig.~\ref{fnh} shows the evolution of N/H, the nitrogen to hydrogen 
ratios, at the surface of our rotating models with $v_{\rm ini}$ = 300 km s$^{-1}$. The ratios
are normalized to their initial value.
The evolutions of the N/H ratio at the surfaces of the non--rotating and rotating stars
are qualitatively similar to the behaviours 
described above for the N/C ratio.
Let us simply say here that
in the plane of Fig.~\ref{fnh}, the non--rotating 
models predict that stars are either on the 
horizontal line defined by $\log  {\rm N/C}=0$ or 
along the line $\log  {\rm N/C}=0.5$--0.6, 
which corresponds to the
position of the non--rotating stars having undergone the first dredge--up.
No stars are predicted outside these two regions. 

Obviously, the observed values obtained by Venn (\cite{Venn99})
 for A and F supergiants in the SMC
are not concentrated along the horizontal lines
predicted by the non--rotating models. Two observed supergiants 
show nitrogen enhancements well below the values predicted by 
the first dredge--up in non--rotating models. Therefore these cannot be on a blue loop. Instead, they
are probably
on their way from the MS to the red 
giant branch and have undergone some mixing
in the early stage of their evolution.
A-- and F--type supergiants observed by Venn (\cite{Venn99})
are  also observed well above the line 
corresponding to the first dredge--up of non--rotating models. 
Again, this is an indication of an extra mixing process
active in massive stars.

Previous studies have shown that for galactic blue supergiants, 
rotation appears as 
a very promising process to account for the observed surface enrichments (see Heger \& Langer \cite{HL00}; paper V;
Meynet \cite{Mey01}). Can we say the same at low metallicity ?
On Fig.~\ref{fnh}, the dotted and the dashed lines define
the evolution between the end of the
MS phase and the stage corresponding to $\log T_{\rm eff}$ = 3.9.
For stars more massive than about 15 M$_\odot$, the N--excesses
do not change very much during this phase.
For the stars that
 present a blue loop episode (even a ``partial blue loop'' episode like
the 12 M$_\odot$ model),
important enrichments occur before the star settles into the red
supergiant stage.

If all the stars were rotating with an initial velocity of 300 km s$^{-1}$, these models
would predict that the observed A--type supergiants would be around the dashed line in Fig.~\ref{fnh}.
For lower initial velocities the N--enrichments
can be
everywhere
between the horizontal line ``$\log {\rm N/C}=0$ '' and the dashed line. 
For higher initial rotational velocities, the N--enrichments would be higher than the dashed line.
Some very N--rich stars at high luminosities (such as the 
N--rich stars with
N/H $>$ 10 and $\log L$/L$_\odot$ $>$ $\sim$ 4.2) could
be accounted for either by stars with very high initial rotation
now on their redward tracks, or  by very rapid rotators
which, after a red supergiant phase, are going  back to the blue.
Thus we can draw the same conclusions as the ones
 deduced at solar metallicity, namely
that there is observational evidences
for an extra mixing process active in massive stars, and that
rotational mixing appears as a very likely
 process to drive the extra mixing
necessary to account for the observations. 

Moreover,
the higher enrichments obtained at  metallicity lower than  solar
is  in agreement
with the observations by Venn (\cite{ve95, Venn99}). 
Comparing
the range of the N/H values measured at the
 surface of A--type supergiants in the Galaxy
and in the SMC, she obtained that in the SMC, the N/H values cover an interval three times
greater than in the Galaxy. At $\log T_{\rm eff}$ = 3.9
the N/H ratio at the surface of the 20 M$_\odot$ model at $Z$ = 0.004 with $v_{\rm ini}$ = 300 km s$^{-1}$ is two times
greater than the ratio at the surface of the similar model
 at solar metallicity. For the N/C ratio the enhancement
amounts to a factor 2.5,
as can be seen from Fig.~\ref{fnc20}.
Thus, even if the distribution of the rotational velocities are the same at both metallicities, one
expects that the low metallicity stars are relatively
 richer in nitrogen. If there are initially more fast  rotating stars 
at low metallicity, this will reinforce this trend.

Thus rotation not only can account for the observations of N--rich blue supergiants at both solar and SMC
metallicity, but it might also account for the fact that at low metallicity the maximum enrichments observed
are greater than at solar metallicity. Finally, 
we point out that in the present models we have found no primary
nitrogen produced. This is not a difficulty, because the evidence
for the existence of some primary nitrogen concern metallicities
still lower than that of the SMC (cf. Henry et al. \cite{Henry00}).
 An interesting
question for future models is to see whether massive star models
 with rotation
at much lower $Z$ produce any primary nitrogen.

\section{Conclusions}

There are 3 main conclusions of this work:

--1.   The ratios $\Omega/\Omega_{\mathrm{crit}}$ 
of the angular velocity to the break--up angular velocity
grow quite a lot during the evolution of
massive stars at low $Z$, contrary to the case  of massive stars
at solar metallicity. This implies that at lower $Z$ a
larger fraction of massive stars is close to the
break--up velocity. This effect results 
from the lower mass loss rates  and from the significant 
outwards transport of angular momentum by meridional
circulation. This is very consistent with
the observational results by Maeder, Grebel and Mermilliod
(\cite{maegremer}), who observed a larger fraction of Be
stars, when passing from Galaxy, to LMC and SMC.
The question arises  whether  the initial distributions
of the stellar rotation in galaxies of different $Z$ are different.
Some arguments in the quoted paper
suggest that  it might  be the case, but this is still very
 uncertain. However, we show here that even if the distribution of 
$v \sin i$ is not different for massive stars with lower $Z$,
the net result of the evolution on the MS leads to
much faster rotational velocities at the end of the MS phase,
when the metallicity is lower.

--2. Due mainly  to the additional helium brought near the H--burning
shell by rotational mixing and the larger He--core, which both
lead to a less efficient H--burning shell and a smaller associated 
convective zone, the stellar radius of rotating stars
is permitted to inflate during the He--burning phase.
These models account for the formation of numerous
red supergiants at low $Z$, with a blue to red supergiant ratio
$B/R$ consistent with the observations in the SMC cluster NGC 330.

--3. While the  standard models with mass loss usually predict no
N/C  enrichments on the main sequence and in the B-- to F--type
supergiants, unless  they have experienced convective dredge--up 
in red supergiants, the models with rotation predict progressive
N/C  and N/O enrichments along the evolutionary tracks. The predicted
N excesses are in general larger at larger stellar luminosities. 
Also, higher relative excesses of N are predicted at lower metallicity
by the present models.
The comparisons
with the observations by Venn (\cite{Venn98, Venn99})  of
A--type supergiants  show a good agreement for  the size of
the N--enrichments, as well as  for the fact that
 they are larger at lower $Z$.

The major interrogation for the future lies in the line
of the first conclusion. Is rotation more and more important
as we go to  massive stars of lower $Z$ ? This is a question
for which great observational efforts  need to be made.
  If the answer is yes, for whatever reason (star formation
or effect of MS evolution), this will imply that rotation
is a dominant effect and  has important consequences
for the interpretation of the spectra of high redshift galaxies
and for the early nucleosynthesis.

    

\end{document}